# On the Trade-offs between Coverage Radius, Altitude and Beamwidth for Practical UAV Deployments

Haneya Naeem Qureshi, *Student Member, IEEE* and Ali Imran, *Senior Member, IEEE*

***Abstract*—Current studies on Unmanned Aerial Vehicle (UAV) based cellular deployment consider UAVs as aerial base stations for air-to-ground communication. However, they analyze UAV coverage radius and altitude interplay while omitting or over-simplifying an important aspect of UAV deployment, i.e., effect of a realistic antenna pattern. This paper addresses the UAV deployment problem while using a realistic 3D directional antenna model. New trade-offs between UAV design space dimensions are revealed and analyzed in different scenarios. The sensitivity of coverage area to both antenna beamwidth and height is compared. The analysis is extended to multiple UAVs and a new packing scheme is proposed for multiple UAVs coverage that offers several advantages compared to prior approaches.**

***Index Terms*—Unmanned aerial vehicles, aerial base stations, air-to-ground communication, 3D antenna model, antenna beamwidth**

## I. Introduction

The demand for more diverse, flexible, accessible and resilient broadband service with higher capacity and coverage is on the rise. Some of these requirements can be accomplished with UAVs acting as aerial base stations. This is because of the several advantages UAV based communication offers such as higher likelihood of line-of-sight (LoS) path and less scatter and signal absorption as compared to terrestrial systems [1], [2]. Moreover, the demand for increase in capacity is leading towards deployment of small cells in terrestrial networks, resulting in the need for higher cell counts, leading to far larger number of ground sites. This makes the goal of attaining seamless coverage over a wide geographical area through terrestrial systems unfeasible due to limited availability of suitable sites and local regulations. This challenge is likely to aggravate with advent of even smaller mmWave cells offering even more sporadic coverage [3].

Similarly, satellite networks have their own limitations such as high latency, high propagation loss, limited orbit space and high launching costs [4]. On the contrary, UAVs can be deployed quickly with much more flexibility to move from one point to another which is a desirable feature for rapid, on-demand or emergency communications [5]-[7].

UAVs can thus be seen as potential enablers to meet the several challenges of next generation wireless systems by either functioning as complementary architecture with already existing cellular networks to compensate for cell overload during peak times and emergency situations [8],[9] or by serving as stand-alone architecture to provide new infrastructure, especially in remote areas [10]-[12]. In this domain, a new hybrid network architecture for cellular systems by leveraging the use of UAVs for data offloading is proposed in [9], [13]. Another significant application of UAVs is in the emerging Internet of Things (IoT) technology [14] and wireless sensor networks [15], where low altitude UAVs can provide a means to collect the IoT data from devices with limited transmit power and transmit it to their intended receivers.

However, in order to fully reap the benefits of UAV based communication, optimal design of UAVs deployment parameters is of fundamental importance. In this paper, we address the UAV deployment problem by analyzing the trade-offs between key system design parameters such as height, antenna beamwidth, and number of UAVs. By leveraging a more realistic model compared to prior studies on the topic, our analysis reveals several new insights and trade-offs between the design parameters that remain unexplored in existing studies.

### A. Related Work

Several studies have recently addressed UAV deployment for different service requirements, mostly using altitude, transmission power and number of UAVs as the only three deployment parameters. For example, authors in [16] investigate the maximum coverage and optimal altitude assuming one UAV with no interference. The optimal altitude is estimated as a function of maximum allowed path loss and statistical parameters of urban environment. However, this work is limited to a single UAV while using mean value of shadowing (rather than its random behavior) and altitude as the only optimization parameter to control coverage. In [17], authors determine optimal height for maximum coverage for a single UAV based on coverage probability and the information rate of users on the ground at a particular UAV altitude.

Authors in [18] extend the work in [16]-[17] to two UAVs, with and without interference. Based on the path loss models in [16] and [19], optimal altitude is reported in [18] for both maximum coverage and minimum required transmit power. Continuing to analyze altitude versus coverage radius relationship, in [20], the same team of authors address the deployment problem with coexistence between UAV and under laid Device-to-Device (D2D) communication networks.

Apart from coverage area, other performance indicators are also affected by changes in UAV height, such as carrier to interference ratio and handovers. Focusing on mm-wave band, authors in [21] investigate coverage versus carrier to interference ratio patterns using an antenna pattern approximated by a cosine function raised to a power. Building upon the work in [21], the effect of lateral displacement of a UAV on interference and handovers is studied in [22]. Authors in [23] measure Receive Signal Strength Indicator (RSSI) for three UAV based cellular networks using following models:

Okumura-Hata, COST-Hata, and COST Walfish-Ikegami. It is reported that signal strengths decrease faster with increase in altitude. However, this work considers UAVs up to an altitude of 500m because of their path loss models constraints. Focusing on just the altitude as a deployment parameter, study in [24] investigates the altitude estimation of UAVs from a more practical perspective using measurements of the polarization of magnetic field of low-frequency radio signals. Authors in [25] estimate the relative attitude between two communicating UAV nodes where the nodes are equipped with MIMO antenna arrays with diverse polarization.

Other works that study UAV deployment from the perspective of optimal altitude and coverage radius include [26]-[28]. The proposed algorithm in [26] finds the optimal altitude of UAV based on the desired radius of coverage in real time. Another deployment model is considered in [27], in which the coverage area is calculated numerically by considering the altitude of UAV and the location of both UAV and users in the horizontal dimension. Results show that the size of the coverage area is affected by the environment. Ideas introduced in [27] are further elaborated in [28] leading to the conclusion that larger buildings require a higher UAV altitude. Additionally, [28] addresses network coverage through a cognitive relay node network model with a goal to enhance the performance of standard relay nodes.

Optimal trajectory designs are studied in [29] by considering a constant UAV height. Joint optimization of UAV's trajectory, as well as the bandwidth allocation and user partitioning between the UAV and ground base stations is analyzed in [9] and [13]. The authors in these studies aim to maximize the minimum throughput of all mobile terminals in the cell and consider orthogonal spectrum sharing between the UAV and ground base stations. The framework is extended to the spectrum reuse case and results show that the proposed hybrid network with optimized spectrum sharing and cyclical multiple access design significantly improves the spatial throughput over terrestrial networks, while the spectrum reuse scheme can provide further throughput gains compared to orthogonal spectrum sharing. Other UAV trajectory designs are considered in [30]-[32]. Authors in [30] aim to design the UAV trajectory to minimize its mission completion time, while ensuring that each ground station successfully recovers the file with a desired high probability. Optimization of multiuser communication scheduling and association jointly with the UAV's trajectory and power control is addressed in [31]. Two other practical types of UAV trajectories, namely circular flight and straight flight are considered in [32] in order to characterize the energy trade-off in ground-to-UAV communication. Another interesting trade-off between throughput and delay from the perspective of trajectory optimization is studied in [33]. In this work, a new cyclical multiple access scheme is proposed to schedule the communications between the UAV and ground terminals in a cyclical time-division manner based on the flying UAV's position. Under this scheme, the authors in [33] reveal a fundamental trade-off between throughput and access delay.

UAV based relay network optimizations are studied in [34]-[36]. Compared with conventional static relaying, authors in [36] consider mobile relaying, which offers a new degree of freedom for performance enhancement via careful relay trajectory design. The authors in this work show that by optimizing the trajectory of the relay and power allocations adaptive to its induced channel variation, mobile relaying is able to achieve significant throughput gains over the conventional static relaying. Authors in [37] consider the impact of antenna power roll-off in 3G networks for fixed platform height and propose a gain adjustment strategy for circular beam antennas.

However, none of the aforementioned studies [16]-[37] consider the impact of antenna gain pattern on the coverage versus height trade-off. One recent study that takes into account effect of directional antenna considers joint altitude and beamwidth optimization for UAV-enabled multiuser communications [38]. In this study, users are partitioned into disjoint clusters and the UAV sequentially serves all clusters by hovering above the cluster centers one by one. In [38], the authors consider three communication models: downlink multicasting, downlink broadcasting and uplink multiple access. One distinguishing feature of [38] is the conclusion that optimal beamwidth and height critically depend on the communication model considered. However, this study uses a step-wise antenna gain model for analytical tractability and line-of-sight propagation conditions. Another recent study [39] that does consider the effect of antenna also uses a step-wise antenna gain model with only two possible values of antenna gain. Analysis incorporating realistic antenna model has become more important since several studies are already considering implementations of directional antenna in UAV-based cellular systems [40], such as smart WiFi directional antennas with servo motors [41]. While the UAV deployment problem has been investigated in a large number of recent studies as discussed above, to the best of authors' knowledge, this is the first paper to study the optimization of UAV deployment design parameters while using a realistic 3D directional antenna model in the system. The analysis presented in this paper shows that the use of a realistic antenna pattern makes a trend shifting difference in the height versus coverage trade-off and adds a new dimension of beamwidth to the UAV deployment design space that remains unexamined in earlier studies.

### B. Contributions and Organization

The contributions and organization of this paper can be summarized as follows:

- We develop a mathematical framework for UAV deployment design while incorporating a realistic model for a practical directional antenna. Current studies on UAV deployment either ignore the effect of 3D directional antenna [16]-[35] or consider an over-simplified model for antenna gain [38][39]. Therefore, UAV deployment analysis presented in such studies, yields results on optimal height, coverage radius and number of UAVs that may not hold for real UAV deployments with practical directional antenna. We address this problem by using 3GPP defined 3D parabolic antenna pattern whose gain is realistically dependent on not only beamwidth but also three dimensional elevation angle. (Section II and III-A)

- We derive analytical expressions for coverage characterized by received signal strength (RSS) as a function of height, beamwidth and coverage radius. (Section II and III-B)
- We present a mathematical framework to quantitatively analyze trade-offs among the following parameters: (i) cell radius versus beamwidth for varying heights, (ii) cell radius versus height for different beamwidths and (iii) beamwidth versus height for different coverage radii. To the best of authors' knowledge, this paper is first to investigate this interesting interplay among the five key factors that define UAV based coverage design space: antenna beamwidth and angular distance dependent gain, elevation angle dependent probability of line of sight, shadowing, free space path loss and height. (Section IV-A)
- We investigate the impact of key UAV design parameters on RSS and validate our derived expressions for probability density function (PDF) of RSS through simulations. (Section IV-B, IV-C and IV-D)
- The proposed framework is extended to a range of frequencies and environments. (Section IV-C)
- Prior works on UAV deployment design [16], [17] [39] use UAV altitude as the only optimization parameter to control coverage. Contrary to the findings from these prior studies, based on our joint analysis of effect of beamwidth and altitude on coverage, we show that:
  - There exists an optimal beamwidth for given height for maximum coverage radius and vice versa. We also derive an expression for determining optimal beamwidth/height for desired coverage radius.
  - Antenna beamwidth is a more practical design parameter to control coverage instead of UAV altitude. This is concluded by performing comparative analysis of the two by quantifying the sensitivity of coverage to both height and beamwidth. (Section IV-E)
  - Contrary to what has been assumed implicitly or explicitly in prior studies, UAV altitude can not be optimized independent of antenna beamwidth. In fact, both parameters need to be optimized in tandem with each other to plan true coverage.
- Coverage probability patterns with varying tilt angles and asymmetrical beamwidths are presented in Section IV-F, which highlight the capability of our derived equations and the underlying system model to extend the analysis to a wide range of scenarios, such as non-zero tilt angle and asymmetrical beamwidths.
- We also extend the analysis to multiple UAVs. Some recent studies have leveraged circle packing theory to determine the number of UAVs needed to cover a given area [39]. However, this approach has two caveats: 1) It leaves significant coverage holes when two or more UAVs are used to cover an area. 2) The number of UAVs increase dramatically with increase in required coverage probability. To circumvent the problems posed by circle packing theory, we propose use of hexagonal packing and compare our results with that obtained by circle packing. This comparison identifies several further advantages of proposed approach. (Section IV-G)
- Continuing our analysis on multiple UAVs, we determine the optimal beamwidth for different number of UAVs that yields maximum total coverage for a target geographical area (Section IV-G). Results show that proposed multi-UAV deployment framework can meet same coverage requirements with less infrastructure (number of UAVs) compared to existing model [39].

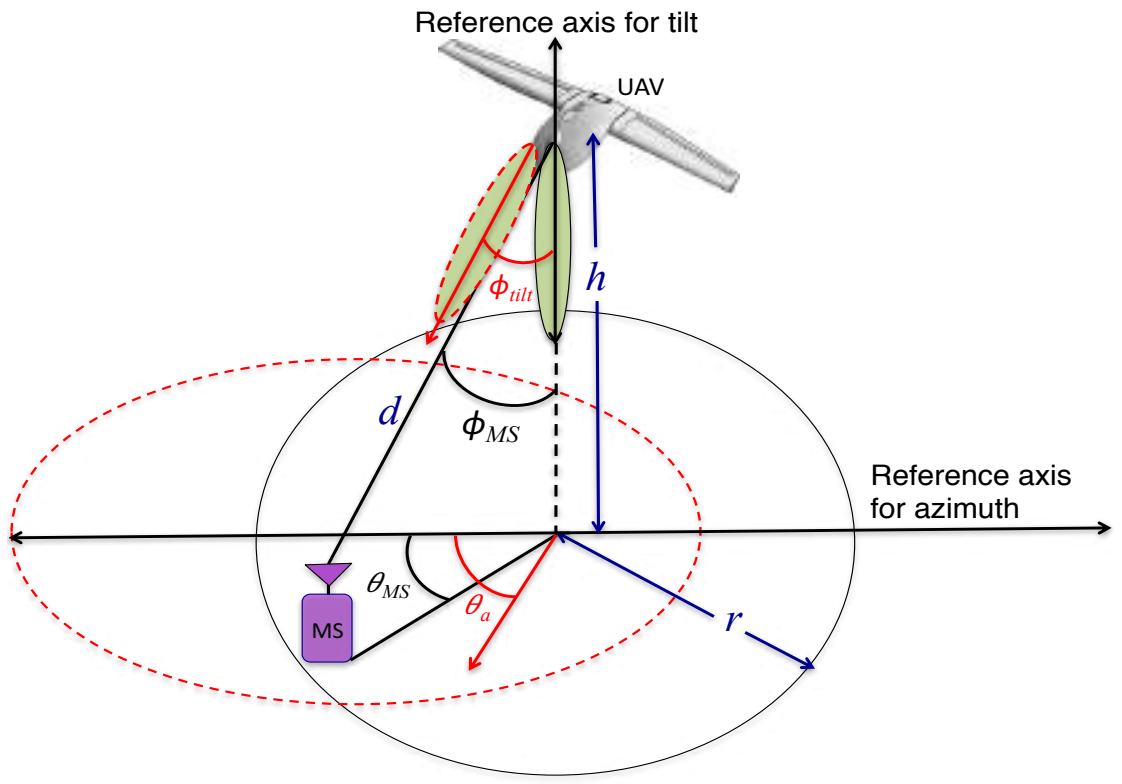

**Fig. 1:** System model.

The key findings of the paper are concluded in Section V.

## II. System Model

We consider a system model illustrated in Fig. 1. The UAV resides at a height $h$ and projects a cell with coverage radius $r$ when $\phi_{tilt} = 0^o$. We define UAV coverage area as a set of points in circle of radius $r$, where a mobile station (MS) experiences a RSS, $S_r$ above a threshold, $\gamma$. Here $r$ is measured from the projection of UAV on ground. In Fig. 1, $\phi_{tilt}$ is the tilt angle in degrees of the antenna mounted on UAV, $\phi_{MS}$ is the vertical angle in degrees from the reference axis (for tilt) to the MS. $\theta_a$ is the angle of orientation of the antenna with respect to horizontal reference axis i.e., positive x-axis and $\theta_{MS}$ is the angular distance of MS from the horizontal reference axis.

Utilizing the geometry in Fig. 1, the perceived antenna gain from the UAV using a three dimensional antenna model recommended by 3GPP [42], at the location of MS can be represented as in (1). Here $B_\theta$ and $B_\phi$ represent the horizontal half power beamwidth (with respect to $\theta$ direction) and the vertical half power beamwidth (with respect to $\phi$ direction) of the UAV antenna in degrees respectively while $\lambda_\theta$ and $\lambda_\phi$ represent the weighting factors for the beam pattern in both directions respectively. $G_{max}$ and $A_{max}$ denote the maximum antenna gain in dB at the boresight of the antenna and maximum attenuation at the sides and back of boresight respectively. $G_{max}$ can be approximated as $10\log\left(\frac{29000}{B_\phi B_\theta}\right)$ [43].

The air-to-ground channel can be characterized in terms of probabilities of LoS and non-line-of-sight (NLoS) scenarios between the UAV and MS. Prior studies have used channel models proposed in [19], [44]-[45]. The channel models proposed in [19] and [44] are suited to only dense urban

$$G(\phi_{MS},\phi_{tilt},\theta_{MS},\theta_a,B_\phi,B_\theta)=\lambda_\phi\left(G_{\max}-\min\left(12\left(\frac{\phi_{MS}-\phi_{tilt}}{B_\phi}\right)^2,A_{\max}\right)\right)+\lambda_\theta\left(G_{\max}-\min\left(12\left(\frac{\theta_{MS}-\theta_a}{B_\theta}\right)^2,A_{\max}\right)\right) \tag{1}$$

**TABLE I:** Key symbol definitions.

| Sym. | Units | Definition | Sym. | Units | Definition |
|---|---|---|---|---|---|
| $h$ | m | height of UAV | $P_l$ | - | probability of LoS scenario |
| $r$ | m | coverage radius | $P_n$ | - | probability of NLoS scenario |
| $d$ | m | UAV to MS distance | $c$ | m/s | speed of light |
| $X_l$ | dB | RV for location variability in LoS scenario | $X_n$ | dB | RV for location variability in NLoS scenario |
| $\sigma_l$ | dB | standard deviation of $X_l$ | $\sigma_n$ | dB | standard deviation of $X_n$ |
| $\phi_{tilt}$ | $^o$ | tilt angle of antenna | $PL_{max}$ | dB | maximum path loss |
| $\phi_{MS}$ | $^o$ | vertical angle from tilt reference axis to MS | $\theta_a$ | $^o$ | angle of antenna w.r.t horizontal reference axis |
| $f$ | GHz | frequency | $X_s$ | dB | RV shadowing |
| $\theta_{MS}$ | $^o$ | angular distance of MS | $\sigma_{s_n}$ | dB | standard deviation of $S_n$ |
| $B_\phi$ | $^o$ | 3dB beamwidth with respect to $\phi$ direction | $B_\theta$ | $^o$ | 3dB beamwidth with respect to $\theta$ direction |
| $\mu'_n$ | dB | mean of $X'_n$ | $\mu_{sh}$ | dB | mean of $X_s$ |
| $\sigma'_n$ | dB | standard deviation of $X'_n$ | $\sigma_{sh}$ | dB | standard deviation of $X_s$ |
| $\lambda_\phi$ | - | weighting factor for beam pattern in $\phi$ direction | $\lambda_\theta$ | - | weighting factor for beam pattern in $\theta$ direction |
| $R_t$ | km | radius of desired geographical area | $\epsilon$ | - | minimum coverage probability |
| $G$ | dBi | antenna gain | $\gamma$ | dBm | received signal threshold |
| $R_l$ | dBm | received signal strength in LoS scenario | $R_n$ | dBm | received signal strength in NLoS scenario |
| $S_n$ | dBm | product of received signal strength in NLoS scenario and probability of NLoS scenario at any $r$: $S_n = P_nR_n$ | $S_l$ | dBm | product of received signal strength in LoS scenario and probability of LoS scenario at any $r$: $S_l = P_lR_l$ |
| $B$ | $^o$ | beamwidth for circular beam pattern | $G_{max}$ | dB | maximum antenna gain at boresight |
| $S_r$ | dBm | received signal strength at any $r$: $S_r = S_l + S_n$ | $S$ | dBm | received signal strength inside a geographical region |
| $P_{cov}$ | - | coverage probability | $\mu_s$ | - | mean of $S_r$ |
| $A_{\max}$ | dB | maximum antenna attenuation at sides and back of boresight | $\sigma_{s_l}$<br>$\sigma_s$ | dB<br>dB | standard deviation of $S_l$<br>standard deviation of $S_r$ |
| $T$ | dBm | transmit signal | $f_S(s)$ | - | PDF of $S$ |
| $N$ | - | number of UAVs | $f_{S_r}(s_r)$ | - | PDF of $S_r$ |

and typical European cities, respectively. Moreover, channel models presented in [19] and [44] lack measurement based validation. On the other hand, the channel model in [45] not only provides a simulation based data for a diverse range of elevation angles, environments and frequencies, but also has been validated through extensive empirical measurements. Hence, we use the UAV channel model proposed in [45] to estimate the probability of LoS scenario as follows:

$$P_l(\phi_{MS}) = 0.01j - \frac{0.01(j-k)}{1+(\frac{90-\phi_{MS}-l}{m})^n} \tag{2}$$

where $(j, ...n)$ are the set of empirical parameters for different types of environments and are given in Table II. The angle, $90-\phi_{MS} = \tan^{-1}(\frac{h}{r})$ is the angle of elevation of the MS to the UAV. The probability of NLoS scenario is then $1 - P_l(\phi_{MS})$.

In addition to free space path loss, UAV-MS signal faces an elevation angle dependent shadowing. The mean and standard deviation for this shadowing can be modeled as [45]:

$$\mu_{sh} = \frac{p_\mu + (90-\phi_{MS})}{q_\mu + t_\mu(90-\phi_{MS})} \tag{3}$$

$$\sigma_{sh} = \frac{p_\sigma + (90-\phi_{MS})}{q_\sigma + t_\sigma(90-\phi_{MS})} \tag{4}$$

where $p_\mu$, $q_\mu$, $t_\mu$, $p_\sigma$, $q_\sigma$ and $t_\sigma$ are parameters obtained from empirical measurements given in Table III. In Table III, the subscript $v = \{\mu, \sigma\}$, is used to indicate that the parameters are for mean and standard deviation, respectively. The RSS in LoS and NLoS scenarios, as a function of path loss and antenna gain can now be represented as:

$$R_l(\phi_{MS},\phi_{tilt},\theta_{MS},\theta_a,B_\phi,B_\theta,d) = T - 20\log\left(\frac{4\pi fd}{c}\right) + G(\phi_{MS},\phi_{tilt},\theta_{MS},\theta_a,B_\phi,B_\theta) - X_l \tag{5}$$

$$R_n(\phi_{MS},\phi_{tilt},\theta_{MS},\theta_a,B_\phi,B_\theta,d) = T - 20\log\left(\frac{4\pi fd}{c}\right) + G(\phi_{MS},\phi_{tilt},\theta_{MS},\theta_a,B_\phi,B_\theta) - X_n - X_s \tag{6}$$

where $T$ is transmitted power, $c$ is the speed of light, $f$ denotes the frequency, $d$ is the distance between UAV and MS and $X_s$ is shadow fading Gaussian $\mathcal{N}(\mu_{sh}, \sigma_{sh})$ random variable (RV) with mean $\mu_{sh}$ and standard deviation $\sigma_{sh}$. $X_s$ is part of the received signal strength in NLoS scenario only because shadowing is a phenomenon associated exclusively with NLoS scenario due to the presence of obstacles in NLoS scenario which affect wave propagation [45]. For realistic system level modeling of mobile systems, random components in dB, $X_l$ and $X_n$ are added as environment dependent variables in LoS and NLoS scenarios [45]. Note that in LoS scenario, despite having a direct path between the UAV and MS, reflections from scatters in the surrounding of the MS can result in different signal strength at different MS locations even when the MS locations are at same distance from the UAV and have LoS. This location dependent randomness in the received signal in LoS scenario is captured in the form of random variable $X_l$ with log-normal distribution of mean zero. $X_l$ and $X_n$ are therefore, $\mathcal{N}(0,\sigma_l)$ and $\mathcal{N}(0,\sigma_n)$ RVs, where $\sigma_l$ and $\sigma_n$ denote the standard deviations, in dB, of $X_l$ and $X_n$ respectively.

**TABLE II:** Environment dependent parameters for $P_l$.

| | Suburban | Highrise urban |
|---|---|---|
| $j$ | 101.6 | 352.0 |
| $k$ | 0 | -1.37 |
| $l$ | 0 | -53 |
| $m$ | 3.25 | 173.8 |
| $n$ | 1.241 | 4.670 |

**TABLE III:** Frequency dependent parameters for shadowing.

| $f$ (GHz) | | $p_v$ | $q_v$ | $t_v$ |
|---|---|---|---|---|
| 2.0 | $v=\mu$ | -94.20 | -3.44 | 0.0318 |
| | $v=\sigma$ | -89.55 | -8.87 | 0.0927 |
| 3.5 | $v=\mu$ | -92.90 | -3.14 | 0.0302 |
| | $v=\sigma$ | -89.06 | -8.63 | 0.0921 |
| 5.5 | $v=\mu$ | -92.80 | -2.90 | 0.0285 |
| | $v=\sigma$ | -89.54 | -8.47 | 0.9000 |

## III. UAV Coverage Model Development

### A. Coverage Probability

As described in Section II, we define UAV coverage area as a set of points in circle of radius $r$, where a MS experiences a RSS, $S_r$ above a threshold, $\gamma$. Then, the RSS at the boundary exceeds a certain threshold, $\gamma = T - PL_{max}$ with a probability, $P_{cov} \geq \epsilon$, where $PL_{max}$ is the maximum allowable path loss. We define this coverage probability, $P_{cov}$ as:

$$\begin{aligned} P_{cov} &= \mathrm{P}[S_r \geq \gamma] \\ &= P_l(\phi_{MS})\mathrm{P}[R_l(\phi_{MS},\phi_{tilt},\theta_{MS},\theta_a,B_\phi,B_\theta,d) \geq \gamma] + \\ &\quad P_n(\phi_{MS})\mathrm{P}[R_n(\phi_{MS},\phi_{tilt},\theta_{MS},\theta_a,B_\phi,B_\theta,d) \geq \gamma] \end{aligned} \tag{7}$$

First, we calculate $\mathrm{P}[R_n(\phi_{MS},\phi_{tilt},\theta_{MS},\theta_a,B_\phi,B_\theta,d) \geq \gamma]$ using (6) as in (8)-(10), where (10) is a result of complementary cumulative distribution function of a Gaussian random variable and substituting $\gamma = T - PL_{max}$ in (9) and $G(\phi_{MS},\phi_{tilt},\theta_{MS},\theta_a,B_\phi,B_\theta)$ is defined in (1). Without loss of generality, we assume $X_n$ and $X_s$ to be independent RVs and thus $X_n' = X_n + X_s$ with mean $\mu_n' = \mu_{sh}$ and standard deviation, $\sigma_n' = \sqrt{\sigma_{sh}^2 + \sigma_n^2}$.

Similarly, we derive $\mathrm{P}[R_l(\phi_{MS},\phi_{tilt},\theta_{MS},\theta_a,B_\phi,B_\theta,d) \geq \gamma]$, which yields the following expression:

$$\mathrm{P}\left[R_l\left(\phi_{MS},\phi_{tilt},\theta_{MS},\theta_a,B_\phi,B_\theta,d\right) \geq \gamma\right] = Q\left(\frac{20\log\left(\frac{4\pi f d}{c}\right) - G(\phi_{MS},\phi_{tilt},\theta_{MS},\theta_a,B_\phi,B_\theta) - PL_{max}}{\sigma_l}\right) \tag{11}$$

where $Q(z) = \frac{1}{\sqrt{2\pi}}\int_z^\infty \exp(\frac{-u^2}{2})\,du$.

Moreover, from cell geometry in Fig. 1 (under the assumption: height of MS << $h$ and the UAV is located at coordinates $(x,y,z) = (0,0,h)$, where $x$ and $y$ are the cartesian coordinates of MS on ground), $\phi_{MS}$ and $d$ can be expressed as:

$$\phi_{MS} = \tan^{-1}\left(\frac{\sqrt{x^2+y^2}}{h}\right), \quad d = \sqrt{h^2+x^2+y^2} \tag{12}$$

The probability of coverage at a particular location of user on ground can now be found by substituting (2), (10) and (11) in (7), which yields the expression given in (13).

However, since UAVs are mobile and can rapidly move from one location to another to provide on-demand coverage where needed, for most practical scenarios, a circular coverage pattern is considered, rather than tampering with antenna tilt. A special case of (13), in which the horizontal and vertical antenna beamwidths are symmetric i.e., $B_\phi = B_\theta = B$ and $\phi_{tilt} = 0$ results in circular coverage footprint of the received signal strength on ground. Note that with $\phi_{tilt} = 0$, the azimuth plane becomes perpendicular to the boresight and the second part of (1) is no longer applicable. In this scenario,

$$r = \sqrt{x^2+y^2}, \quad \phi_{MS} = \tan^{-1}\left(\frac{r}{h}\right), \quad d = \sqrt{h^2+r^2} \tag{14}$$

Further, $G_{max}$ can be approximated as $10\log\left(\frac{29000}{B^2}\right)$ [43] and $A_{max}$ can be ignored without impacting the required accuracy of this antenna model that mainly concerns gain on and around the boresight. Applying these simplifications to (1), substituting (1) and (14) in (10)-(11) and then making use of (7) yields the expression for $P_{cov}$ as a function of antenna beamwidth, UAV height and coverage radius in (15).

### B. Received Signal Strength

One way to investigate the RSS on a particular location on ground for a given height and antenna beamwidth for circular coverage pattern is by evaluating the expected value of $S_r$ over the random variables $X_s$, $X_l$ and $X_n$ as follows

$$\begin{aligned} \mathbf{E}[S_r] &\overset{\text{(k)}}{=} P_l\,\mathbf{E}[S_r|\text{LoS}] + P_n\,\mathbf{E}[S_r|\text{NLoS}] \\ &= P_l\,\mathbf{E}[R_l - R_n] + \mathbf{E}[R_n] \\ &= \left(0.01j - 1 - \frac{0.01(j-k)}{1+\left(\frac{\tan^{-1}(\frac{h}{r})-l}{m}\right)^n}\right)\mu_{sh} - 20\log\left(\frac{4\pi f d}{c}\right) \\ &\quad + T - 12\left(\frac{\tan^{-1}\left(\frac{r}{h}\right)-\phi_{tilt}}{B}\right)^2 + 10\log\left(\frac{29000}{B^2}\right) \end{aligned} \tag{16}$$

$$\mathrm{P}\left[R_n\left(\phi_{MS},\phi_{tilt},\theta_{MS},\theta_a,B_\phi,B_\theta,d\right)\geq\gamma\right]=\mathrm{P}\left[T+G(\phi_{MS},\phi_{tilt},\theta_{MS},\theta_a,B_\phi,B_\theta)-20\mathrm{log}\Big(\frac{4\pi fd}{c}\Big)-X_n-X_s\geq\gamma\right] \tag{8}$$

$$=\int_\gamma^\infty\frac{1}{\sqrt{2\pi}\sigma_n'}\exp\left[-\frac{1}{2}\left(\frac{X_n'-\Big(T+G(\phi_{MS},\phi_{tilt},\theta_{MS},\theta_a,B_\phi,B_\theta)-20\mathrm{log}\left(\frac{4\pi fd}{c}\right)-\mu_n'\Big)}{\sigma_n'}\right)^2\right]dX_n' \tag{9}$$

$$=Q\left(\frac{20\mathrm{log}\left(\frac{4\pi fd}{c}\right)+\mu_{sh}-G(\phi_{MS},\phi_{tilt},\theta_{MS},\theta_a,B_\phi,B_\theta)-PL_{max}}{\sqrt{\sigma_{sh}^2+\sigma_n^2}}\right) \tag{10}$$

$$P_{cov}(\phi_{MS},\phi_{tilt},\theta_{MS},\theta_a,d,B_\theta,B_\phi)=\left(0.01j-\frac{0.01(j-k)}{1+\left(\frac{90-\phi_{MS}-l}{m}\right)^n}\right)Q\left(\frac{20\log\left(\frac{4\pi fd}{c}\right)-G(\phi_{MS},\phi_{tilt},\theta_{MS},\theta_a,B_\phi,B_\theta)-PL_{max}}{\sigma_l}\right)+$$
$$\left(1-0.01j+\frac{0.01(j-k)}{1+\left(\frac{90-\phi_{MS}-l}{m}\right)^n}\right)Q\left(\frac{20\mathrm{log}\left(\frac{4\pi fd}{c}\right)+\mu_{sh}-G(\phi_{MS},\phi_{tilt},\theta_{MS},\theta_a,B_\phi,B_\theta)-PL_{max}}{\sqrt{\sigma_{sh}^2+\sigma_n^2}}\right) \tag{13}$$

$$P_{cov}^c(r,h,B)=\left(0.01j-\frac{0.01(j-k)}{1+\left(\frac{\tan^{-1}\left(\frac{h}{r}\right)-l}{m}\right)^n}\right)Q\left(\frac{20\log\left(\frac{4\pi f\sqrt{h^2+r^2}}{c}\right)-10\log\left(\frac{29000}{B^2}\right)+12\left(\frac{\tan^{-1}\left(\frac{r}{h}\right)-\phi_{tilt}}{B}\right)^2-PL_{max}}{\sigma_l}\right)+$$
$$\left(1-0.01j+\frac{0.01(j-k)}{1+\left(\frac{\tan^{-1}\left(\frac{h}{r}\right)-l}{m}\right)^n}\right)Q\left(\frac{20\mathrm{log}\left(\frac{4\pi f\sqrt{h^2+r^2}}{c}\right)+\mu_{sh}-10\log\left(\frac{29000}{B^2}\right)+12\left(\frac{\tan^{-1}\left(\frac{r}{h}\right)-\phi_{tilt}}{B}\right)^2-PL_{max}}{\sqrt{\sigma_{sh}^2+\sigma_n^2}}\right) \tag{15}$$

where (k) is a result of the law of total expectation and $P_l, R_l$ and $R_n$ are functions of $r, h$ and $B$ and defined in (2), (5) and (6) respectively. (16) is therefore obtained by substituting $R_l$ and $R_n$ from (5) and (6) and then making use of the relationship between $P_l(\phi_{MS})$ and $P_n(\phi_{MS})$, i.e., $P_n(\phi_{MS}) = 1 - P_l(\phi_{MS})$.

However, $R_l$ is a random variable due to the random component, $X_l$ and $R_n$ is a random variable due to the random components, $X_n$ and $X_s$. Therefore, RSS, can also be modeled as a random variable. In order to derive an analytical expression for the PDF of RSS at any arbitrary cell location, we express it as:

$$S_r(h,r,B)=S_l(h,r,B)+S_n(h,r,B) \tag{17}$$

where $S_l$ and $S_n$ are independent random variables, given by $S_l(h,r,B) = P_l(\phi_{MS})R_l(h,r,B)$ and $S_n(h,r,B) = P_n(\phi_{MS})R_n(h,r,B)$. Then, in order to derive an analytical expression for PDF of $S_r$, we resort to transformations of random variables and convolution of the PDFs of $S_l$ and $S_n$, resulting in the expression in (18). Complete derivation of (18) is provided in Appendix.

The normalized PDF of received signal strength inside a geographical region (denoted by $S$) by assuming that the UAV resides at coordinates $(x, y, z) = (0, 0, h)$ where $h$ is height of UAV, can then be found as follows:

$$f_S(s)=\frac{1}{A}\iint_A f_{S_r}(s,x,y)dxdy \tag{19}$$

in some geographical region $A$ that lies in the $xy$-plane. The integral in (19) can be solved through numerical methods.

## IV. Numerical Results and Analysis

### A. Trade-offs between coverage radius, beamwidth and height

*1) Coverage Radius vs. Beamwidth:* As the height of UAV increases for a particular MS location, $\phi_{MS}$ in (14) decreases (angle of elevation increases), leading to an increase in probability of LoS link in (2), decrease in shadowing in (4) and increase in free space path loss (as $d$ increases). While the effect of these factors on coverage has been studied in earlier studies [16]-[37], the impact of the fourth factor, antenna gain in conjunction with these three factors remained unexamined. Fig. 2 shows that impact of this fourth factor is so profound that at a given cell radius, it results in a height ($h'$), after which the antenna gain trend with increasing beamwidth reverses. Fig. 2 is plotted utilizing (1) for $\phi_{tilt} = 0$ at $r = 5000$ m under the assumptions stated in Section III. The larger antenna gain at a given $r$ is observed with increased height because, for the same MS location, $\phi_{MS}$ in (14) decreases. For any two beamwidths, $B_1$ and $B_2$, $h'$ is the point of intersection of gain versus height graphs illustrated in Fig. 2 for $r = 5000$ m. $h'$ can be calculated as follows:

$$10\log\left(\frac{29000}{B_1^2}\right)-12\left(\frac{\tan^{-1}\left(\frac{r}{h'}\right)}{B_1}\right)^2=10\log\left(\frac{29000}{B_2^2}\right)-12\left(\frac{\tan^{-1}\left(\frac{r}{h'}\right)}{B_2}\right)^2$$

$$h'=r\left(\tan\sqrt{\frac{5}{3}\log\left(\frac{B_2}{B_1}\right)\left(\frac{B_1^2B_2^2}{B_2^2-B_1^2}\right)}\right)^{-1} \tag{20}$$

The coverage radius versus beamwidth trend for different heights in Fig. 3 can now be analyzed in light of the aforementioned factors. This figure is plotted by utilizing (15) for

$$f_{S_r}(s_r) = \frac{\exp\left[\frac{-\left[s_r-\left(1-0.01j+\frac{0.01(j-k)}{1+\left(\frac{\tan^{-1}(h/r)-l}{m}\right)^n}\right)\left(\frac{p_\mu+\tan^{-1}(h/r)}{q_\mu+t_\mu\tan^{-1}(h/r)}\right)-T+20\log\left(\frac{4\pi f\left(\sqrt{h^2+r^2}\right)}{c}\right)-10\log\left(\frac{29000}{B^2}\right)+12\left(\frac{\tan^{-1}(r/h)}{B}\right)^2\right]^2}{2\sigma_l^2\left(\frac{0.01(j-k)}{1+\left(\frac{\tan^{-1}(h/r)-l}{m}\right)^n}\right)^2+\left(1-\frac{0.01(j-k)}{1+\left(\frac{\tan^{-1}(h/r)-l}{m}\right)^n}\right)^2\left(\sigma_n^2+\left(\frac{p_\sigma+\tan^{-1}(h/r)}{q_\sigma+t_\sigma\tan^{-1}(h/r)}\right)^2\right)}\right]}{\sqrt{2\pi\sigma_l^2\left(\frac{0.01(j-k)}{1+\left(\frac{\tan^{-1}(\frac{h}{r})-l}{m}\right)^n}\right)^2+\left(1-\frac{0.01(j-k)}{1+\left(\frac{\tan^{-1}(\frac{h}{r})-l}{m}\right)^n}\right)^2\left(\sigma_n^2+\left(\frac{p_\sigma+\tan^{-1}\left(\frac{h}{r}\right)}{q_\sigma+t_\sigma\tan^{-1}\left(\frac{h}{r}\right)}\right)^2\right)}} \quad (18)$$

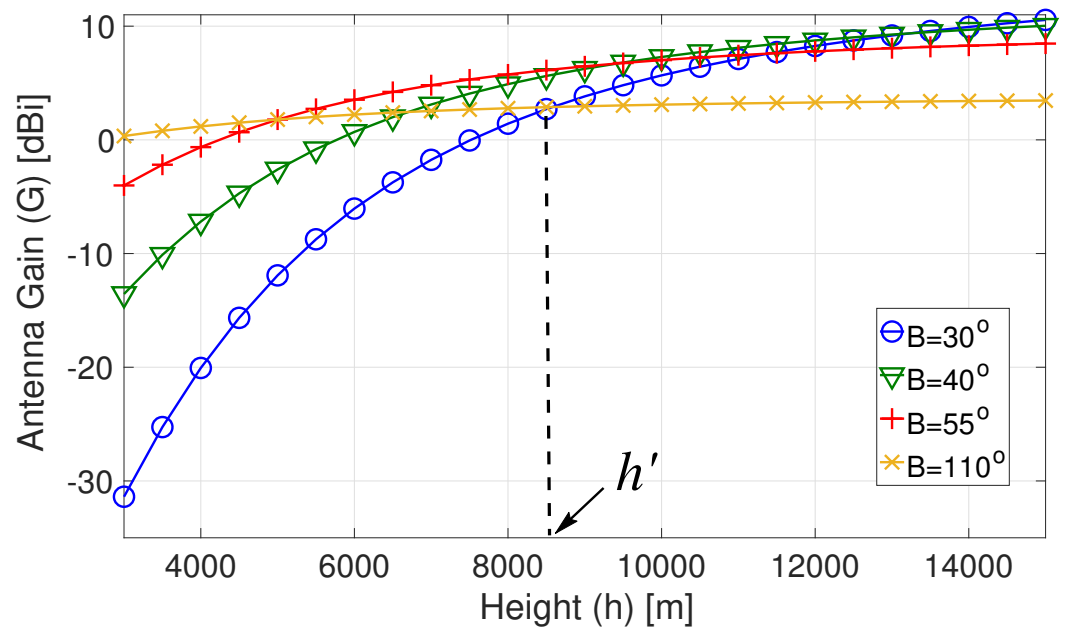

**Fig. 2:** Antenna gain at $r = 5000$ m with varying heights for different beamwidths.

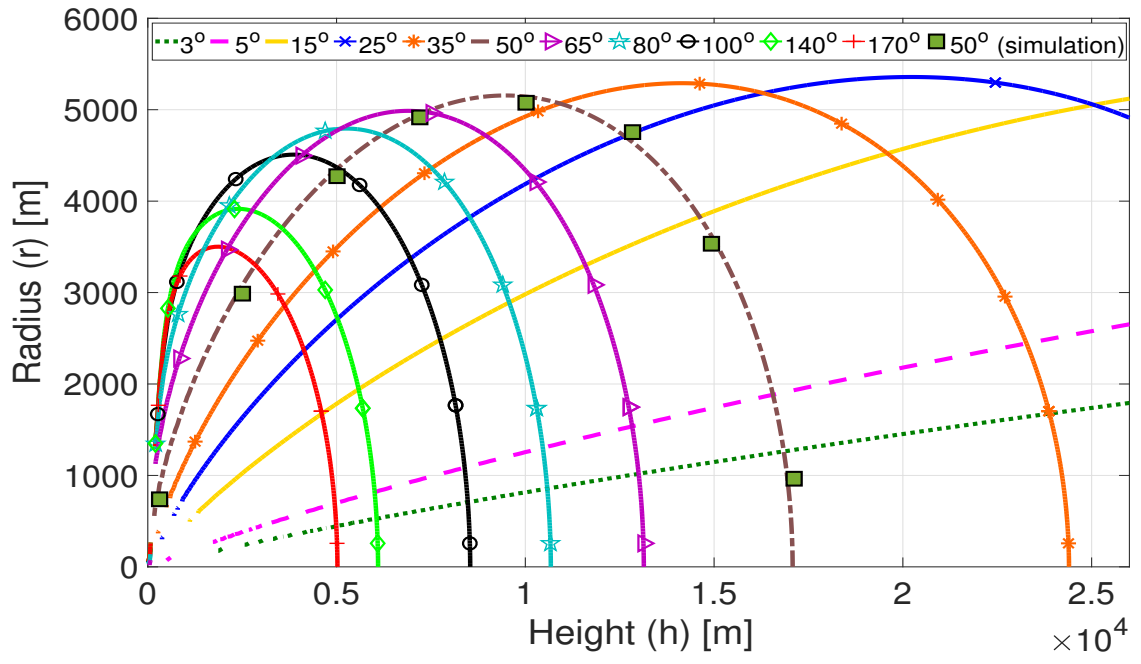

**Fig. 4:** Coverage radius against height for different beamwidths.

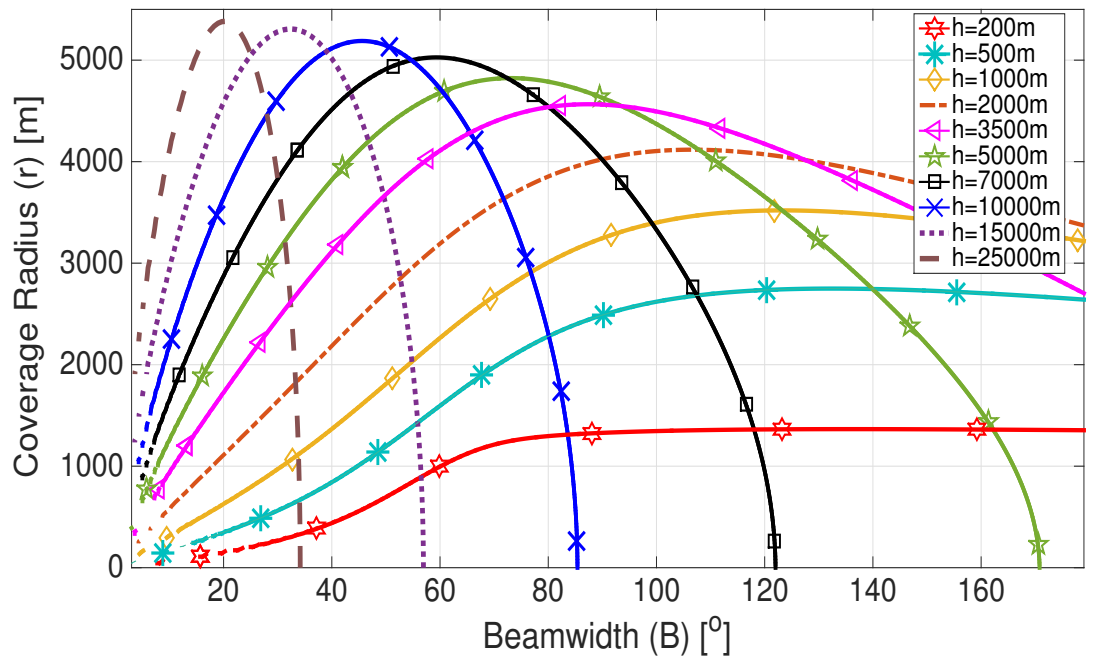

**Fig. 3:** Coverage radius against beamwidth for different heights.

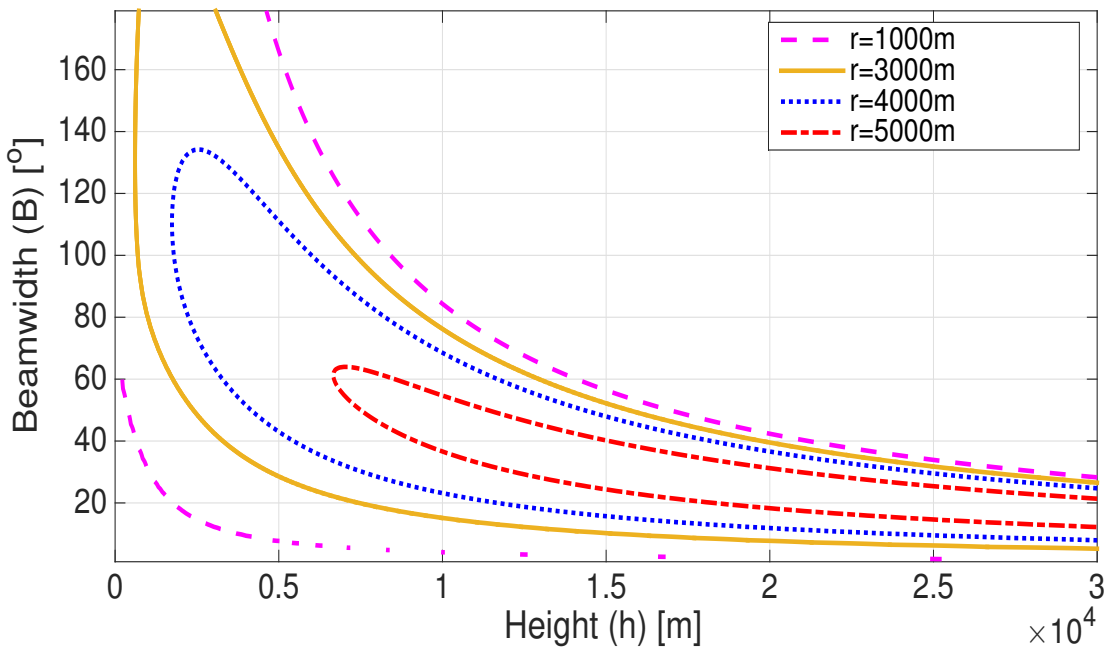

**Fig. 5:** Beamwidth against height for different radii.

$PL_{max}$=115 dB, $\phi_{tilt} = 0^o$, $\epsilon = 0.8$ and $f = 2$ GHz in a suburban environment. The trend shift in Fig. 3 is attributed to the following: as height increases up to 1000 m, the increase in antenna gain and decrease in shadowing offsets the increased free space path loss. As height increases beyond 1000 m, the increase in antenna gain and decrease in shadowing is overshadowed by the increase in free space loss. As a result, the coverage radius increases with beamwidth, approaches to a maximum value and then starts to decrease. Our analysis quantifies this maximum value of coverage radius in relation to antenna gain, for instance, maximum coverage radius of 5000 m at a beamwidth of $55^o$ for $h = 7000$ m in Fig. 3 can be attributed to the occurence of largest antenna gain at a beamwidth of $55^o$ at a height of 7000 m in Fig. 2.

*Our UAV coverage model for the first time shows the existence of optimal beamwidth for given height for maximum coverage radius and vice versa, a trend that remains hidden in UAV coverage models presented in prior studies [16]-[39].*

*2) Coverage Radius vs. Height:* Fig. 4 depicts the relation of coverage radius with height for different beamwidths. Initially, as the height of the UAV increases for small beamwidths, coverage radius also increases continuously. However, as beamwidth increases further, coverage radius versus height curves attain a parabolic shape. This is in contrast to previous studies that suggest monotonic increase of UAV coverage radius with altitude [26] [39]. Thus, our analysis brings forth these new insights as it captures the relative effect of all four factors that impact the coverage radius concurrently: angular distance dependent realistic non-linear antenna model, elevation angle dependent probability of line of sight, shadowing, and a measurement backed path loss model.

*3) Height vs. Beamwidth:* There can be scenarios where height of UAV is subject to changes due to factors beyond the system designer's control such as weather, but the same coverage pattern has to be maintained. Our proposed model provides a mechanism to address such scenarios by characterizing the height versus beamwidth relationship as shown in Fig. 5. For example, if a UAV is deployed to cover $r_f = 4000$ m with $\epsilon = 0.8$ at a height of 10000 m and if its height is changed to 5000 m, the UAV will need to adjust its beamwidth from $70^o$ to either $42^o$ or $110^o$ in order to continue providing the same coverage. Fig. 5 also highlights the importance of optimizing both beamwidth and height in tandem with each other rather than independently as has been the case in prior works [16] and [39].

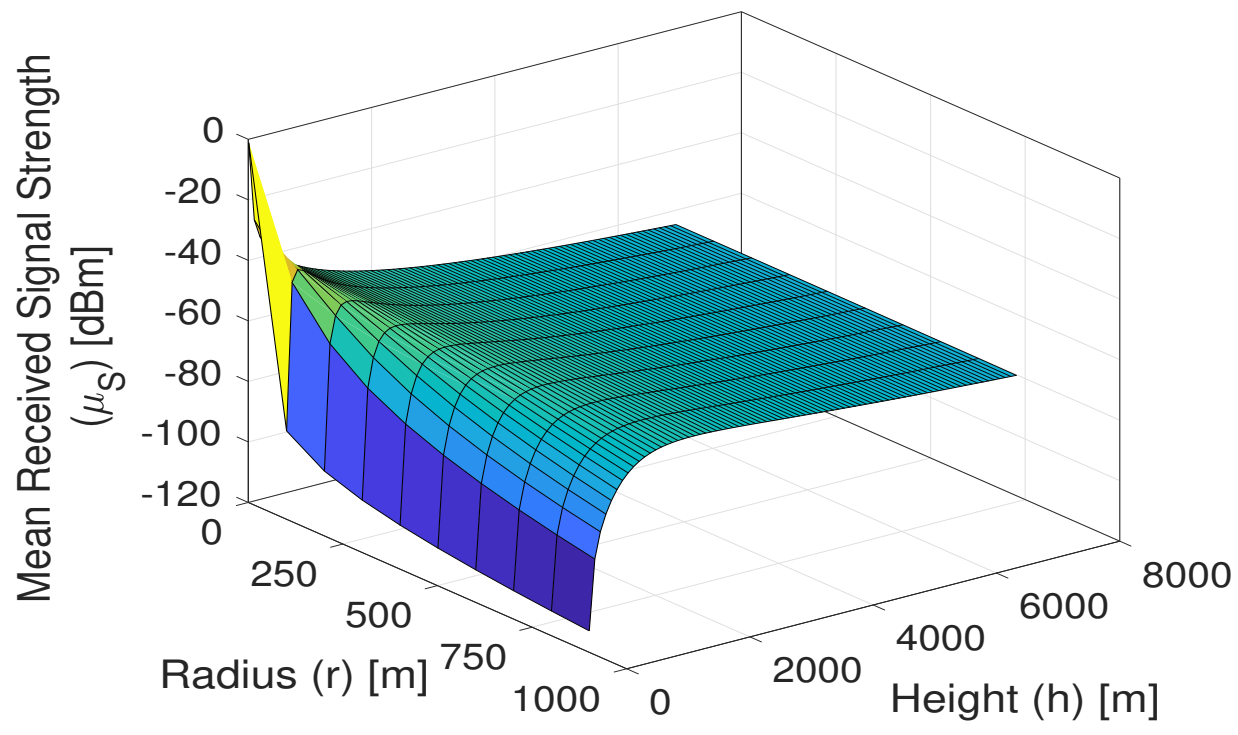

**Fig. 6:** Mean received signal with varying height for $B = 50^o$.

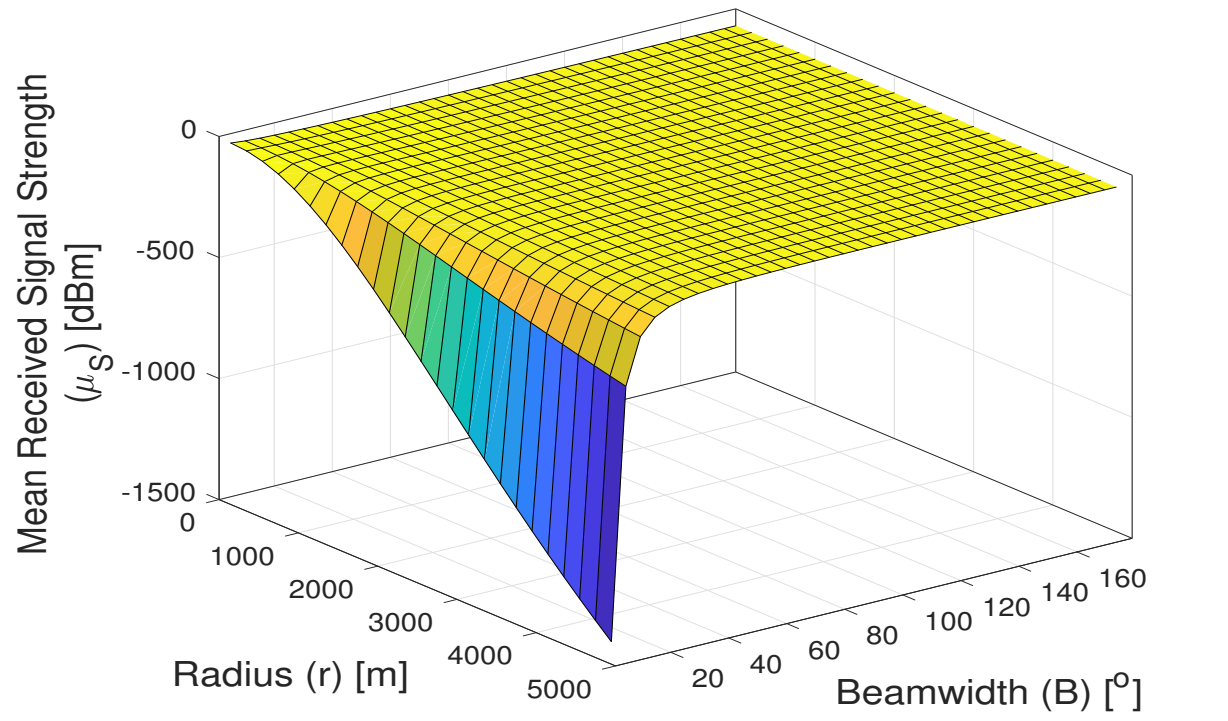

**Fig. 7:** Mean received signal with varying beamwidth at $h = 3500$ m.

## *B. Impact of altitude, beamwidth and radius on RSS*

In order to investigate the behavior of RSS in context with UAV design parameters, Fig. 6 and Fig. 7 illustrate how the mean RSS varies with height and beamwidth with increasing cell radius $r$. In Fig. 6, mean RSS is plotted for $B = 50^o$, $f = 2$ GHz, $P_t = 40$ dBm and $\phi_{tilt} = 0^o$ in a suburban environment. As expected, RSS decreases as $r$ increases for any fixed height. However, for any $r$, RSS initially increases with height, reaches to a peak value and then decreases as height increases further. The scenario in Fig. 7 consists of a UAV deployed at a height of 3500 m. Since a narrow beamwidth can only cover a small coverage area, the decrease in RSS with increasing radius is very rapid at low beamwidths. Hence, the trend of RSS with beamwidth is more clearly depicted in Fig. 8a-8c, which are zoomed-in plots of Fig. 7 for $20^o \leq B < 180^o$. The continuous decrease of coverage radius with increase in beamwidth for $r < 800$ m in Fig. 8a can be attributed to the continuous decrease of antenna gain with radius in Fig. 9 for $r < 800$ m. As we approach closer to 1000m, the trend in antenna gain pattern starts to reverse for different beamwidths, and we observe several intersecting points from $1000 < r < 2500$ in Fig. 9 and hence mean RSS in Fig. 8b attains a parabolic shape with increasing beamwidth in this range. At very large $r$, antenna gain trend reverses completely and now it increases with increasing beamwidths, which is again in line with the RSS trend in Fig. 8c.

Note also that at very large coverage radius, for example, at $r = 4000$ m in Fig. 8c, RSS decreases very sharply from $50^o$ to $30^o$ as compared to decrease in RSS from $120^o$ to $80^o$. This is because at 4000 m in Fig. 9, difference in gain between $50^o$ and $30^o$ is quite high as compared to difference in gain between $120^o$ and $80^o$.

## *C. Analysis for different frequencies and environments*

In Fig. 10, we quantify the trade-off of coverage radius with beamwidth in changing environments and at different frequencies. The figure is plotted for $h = 3000$ m, $PL_{max} = 120$ dB for two extreme environments, suburban and high rise urban. From Fig. 10, we observe that in addition to coverage radius decreasing sharply as frequency increases or environment becomes denser, the beamwidth at which this coverage radius versus beamwidth trend changes is also lower in a more denser environment or at a higher frequency. For example, in high rise urban environment, decrease in radius starts from a beamwidth of as low as $30^o$ at 2.0 GHz, whereas for the same frequency, this decrease does not start until $110^o$ in case of the suburban environment. Similar observations can be made by observing the effect of frequency in the same environment. This is not just because of free space path loss which increases with increasing frequency, but also because of shadowing, which increases at higher frequencies for the same environment [45]. In addition, the impact of frequency on coverage radius reduces as environment becomes denser. These observations could play a valuable role for designing UAV based cellular systems at higher frequencies by utilizing the unused part of higher frequency spectrum such as mmWave. The analytical results are also corroborated with simulation results in Fig. 10.

## *D. Validation of analysis with Monte Carlo Simulations*

The trade-offs between coverage radius, beamwidth and height presented in the preceding sections have already been verified through monte-carlo simulations as shown in Fig. 4 and Fig. 10. Next, we corroborate the results derived for RSS model in (17)-(19) via simulations. Fig. 11 and Fig. 12 show PDF of RSS at two arbitrary points, located at $r = 3000$ m for UAVs deployed at height of $h = 2000$ m and beamwidths of $B = 5^o$ and $50^o$ respectively. The RSS PDF obtained from our derived analytical expression in (18) shows an excellent fit to simulation based RSS data. In Fig. 11, users located at $r = 3000$ m receive extremely low RSS since a small beamwidth of $5^o$ can cover only a small area. In contrast, when beamwidth is increased to $50^o$, the same users start to receive a much better coverage i.e., between -85 to -65 dBm.

We now extend our analysis from RSS at any arbitrary point to RSS inside a geographical area of 8000 m $\times$ 8000 m. Normalized histograms of RSS from the simulation results and analytical PDFs from (19) are in agreement as shown in Fig. 13-14. From Fig. 13-14, we note that not only the range of RSS becomes narrower in a given area, but also the distribution of RSS approaches zero skewed Gaussian with either increasing height or increasing beamwidth.

## *E. Comparison of altitude and beamwidth to control coverage*

Noting that both beamwidth and height can be used for the same purpose of controlling coverage leads us towards comparing both of these parameters by analyzing the sensitivity of coverage radius to each of these parameters.

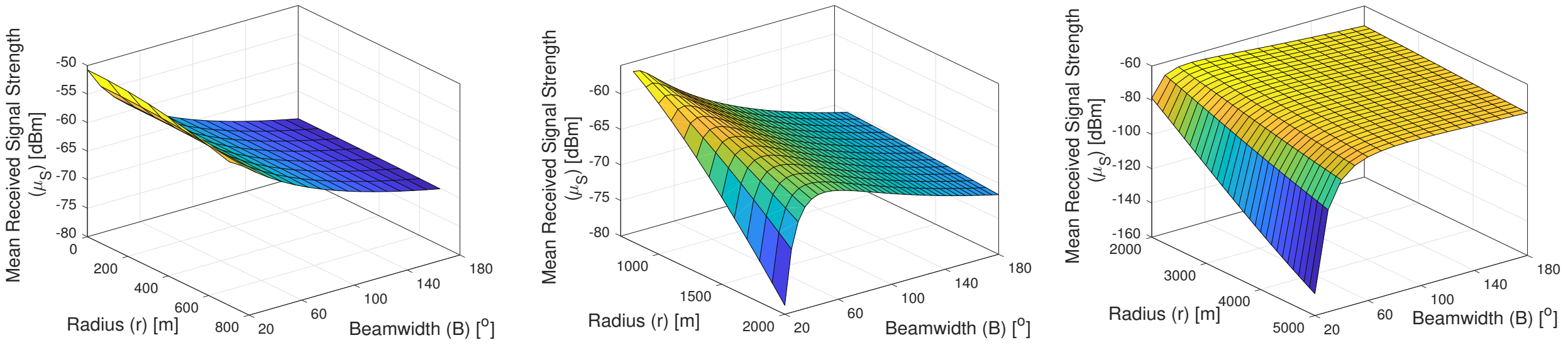

**(a)** 0 m $< r <$ 800 m **(b)** 800 m $< r <$ 2000 m **(c)** 2000 m $< r <$ 5000 m

**Fig. 8:** Mean received signal on ground with varying beamwidths $> 20^o$

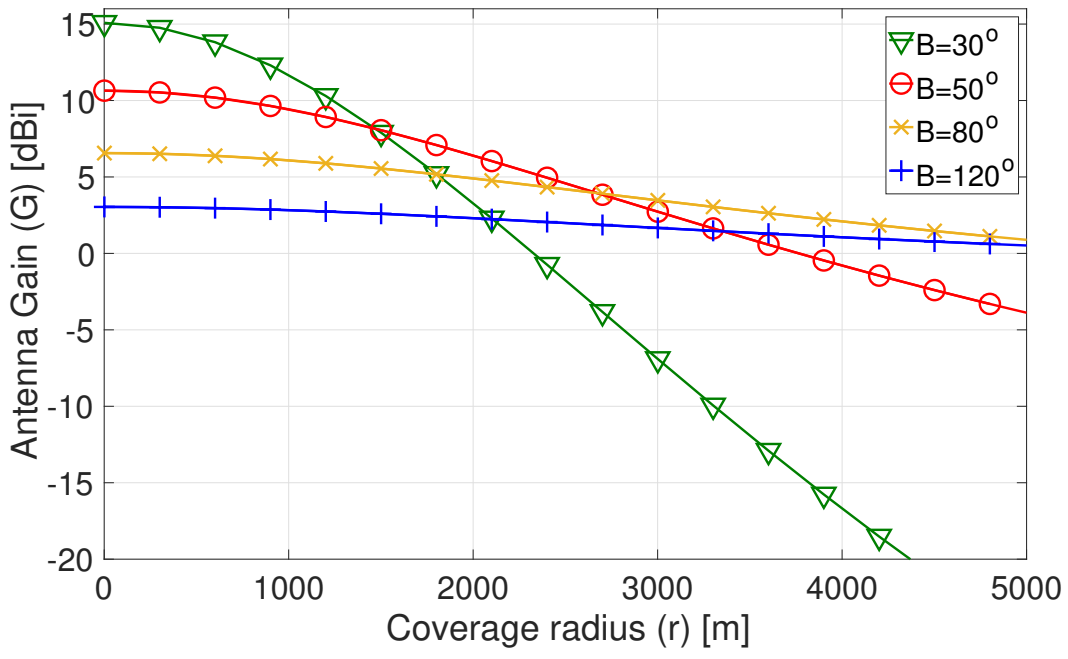

**Fig. 9:** Antenna gain at $h = 3500$m with varying radius for different beamwidths.

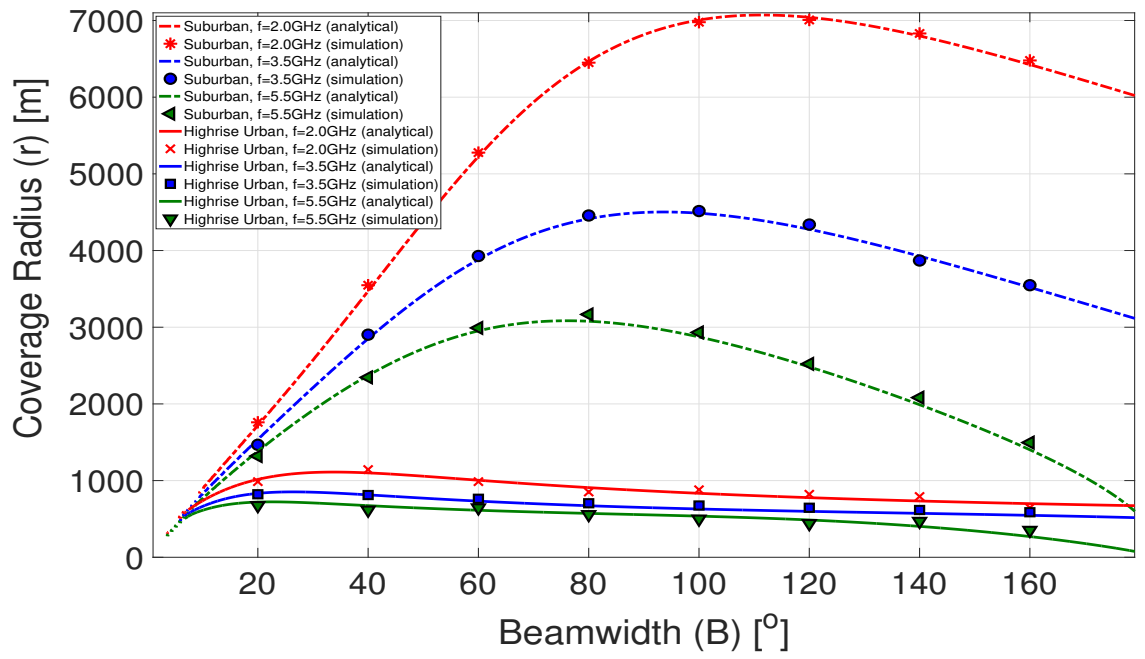

**Fig. 10:** Coverage radius against beamwidth for varying frequency and environment at $h = 3000$ m.

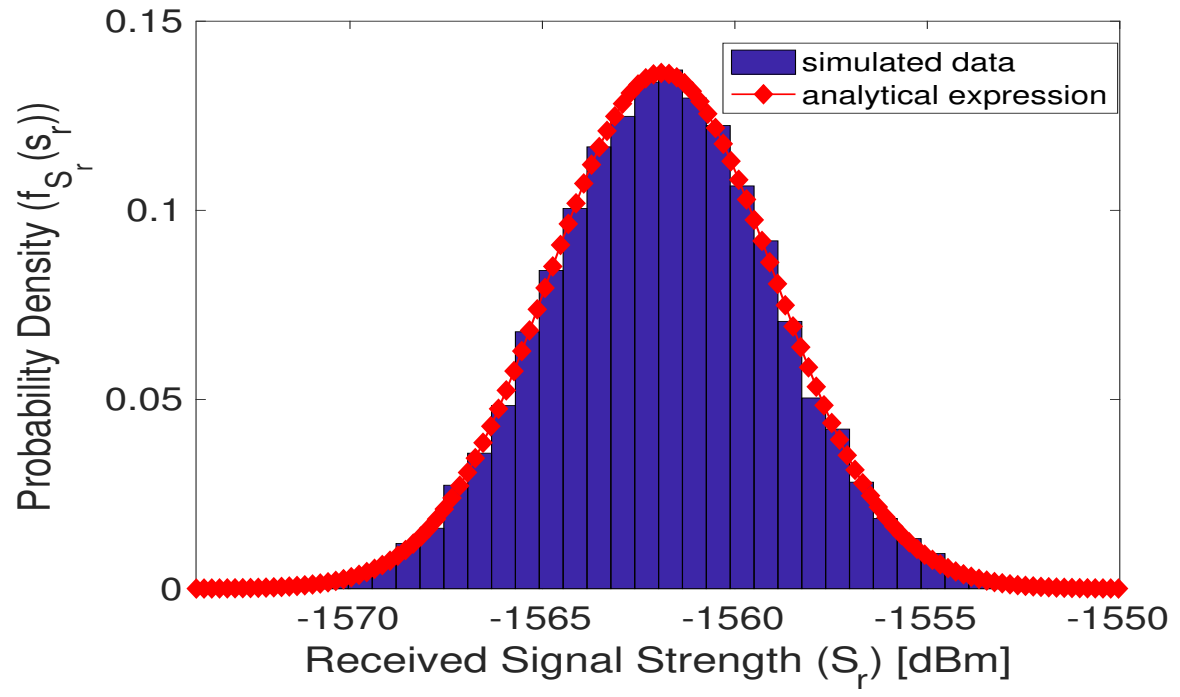

**Fig. 11:** PDF of RSS at $r = 3000$ m, $B = 5^o$ and $h = 2000$ m.

Fig. 15-16 show the gradient of radius with respect to beamwidth obtained by differentiating (15) with respect to beamwidth or determining gradient of the curves in Fig. 3 with respect to beamwidth. By comparing the absolute values of $\Delta r/\Delta B$ in Fig. 15-16, we conclude that the change in coverage radius is most sensitive to very high heights (upto 25000 m).

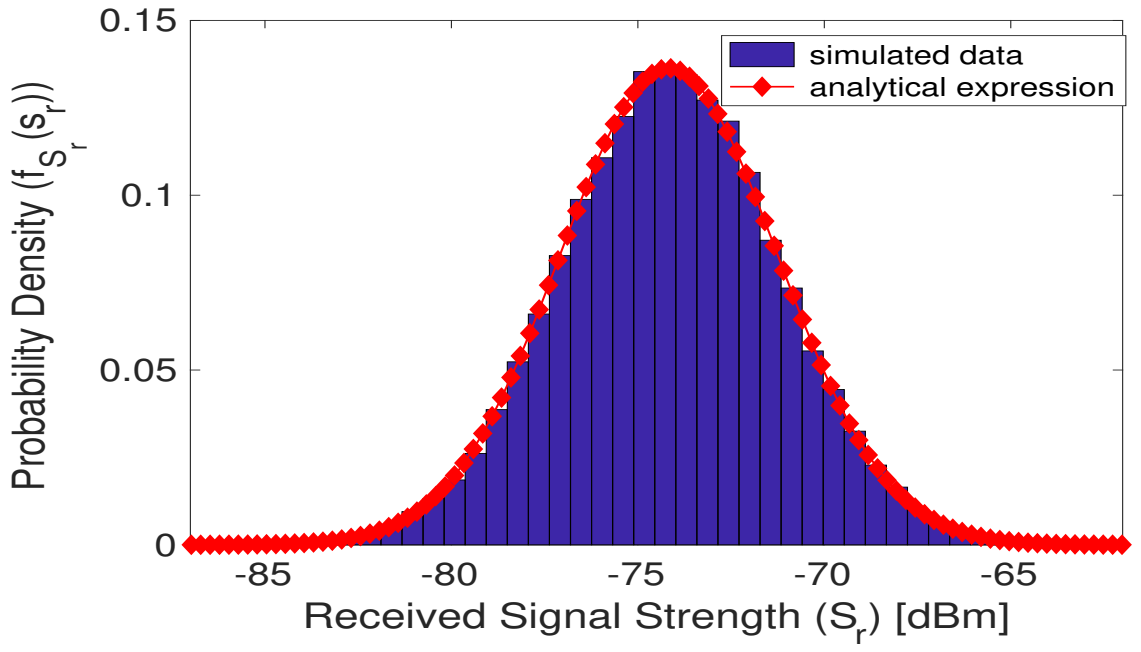

**Fig. 12:** PDF of RSS at $r = 3000$ m, $B = 50^o$ and $h = 2000$ m.

Next, we analyze the rate of change of radius with height ($\Delta r/\Delta h$) in Fig. 17. Unlike Fig.15, the values of derivatives here follow a similar pattern with changing beamwidths except for beamwidths $< 25^o$ where $\Delta r/\Delta h$ is constant. By comparing values of the derivative in Fig. 17 with Fig. 15-16 (both comparisons range from samples of heights from between 0 to 25000 m and beamwidths from $1^o$ to $180^o$), we note that $\max(\Delta r/\Delta B) >> \max(\Delta r/\Delta h)$. Numerically, the maximum value of $\max(\Delta r/\Delta B)$, -2000 is almost 75 times greater than the maximum value of $\max(\Delta r/\Delta h)$, -27.

This indicates that low beamwidths lead to greatest change (decrease) in coverage radius per unit beamwidth as compared to change in radius per unit height. This analysis can be leveraged to choose between the height and beamwidth or design appropriate combination of the two for optimizing coverage.

### *F. Coverage probability with varying tilt angles and asymmetric beamwidths*

Coverage analysis of scenarios with varying tilt angles and asymmetrical beamwidths is presented in this section. In Fig. 18, we compare the probability of coverage using zero antenna tilt with a tilt angle of $10^o$. For non-zero tilt angles, the coverage pattern forms an off-centered ellipse shape rather than a circle centered at the origin. The coverage probability in case of non-zero tilt angle is evaluated by using (13). Fig. 19 shows how the coverage probability changes with different values of $B_\phi$ and $B_\theta$. The effect of changing tilt values is depicted in Fig. 20. It is observed that although a UAV with antenna tilt of $80^o$ covers more area as compared to tilt angle of $20^o$, the maximum coverage probability with $\phi_{tilt} = 80^o$ is reduced by half as compared to $\phi_{tilt} = 20^o$.

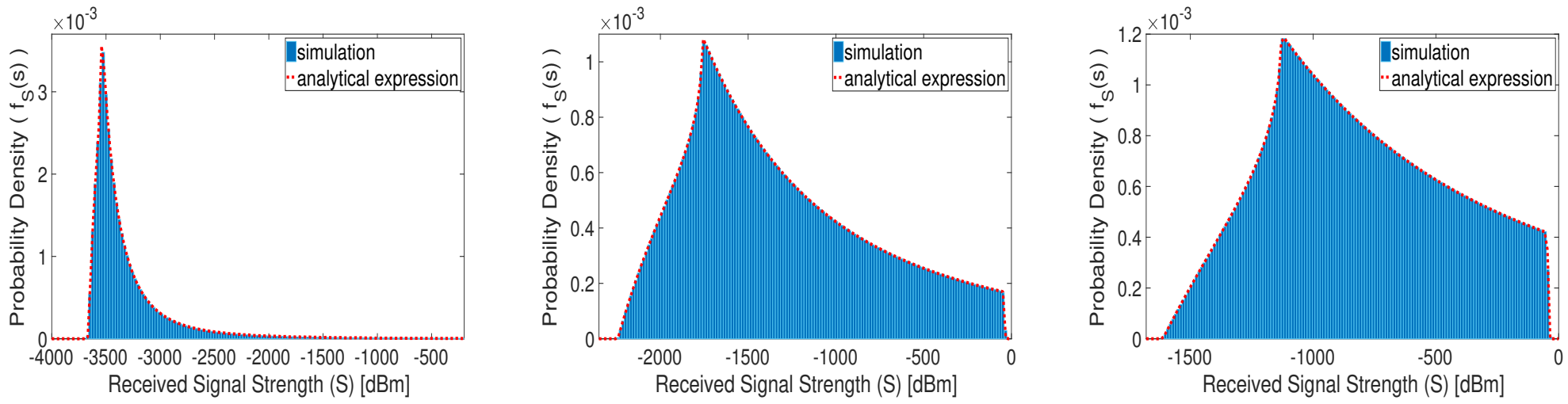

**(a)** $h = 500$ m **(b)** $h = 3500$ m **(c)** $h = 5500$ m

**Fig. 13:** PDF of RSS on ground with changing altitude of UAV for $B = 5^o$

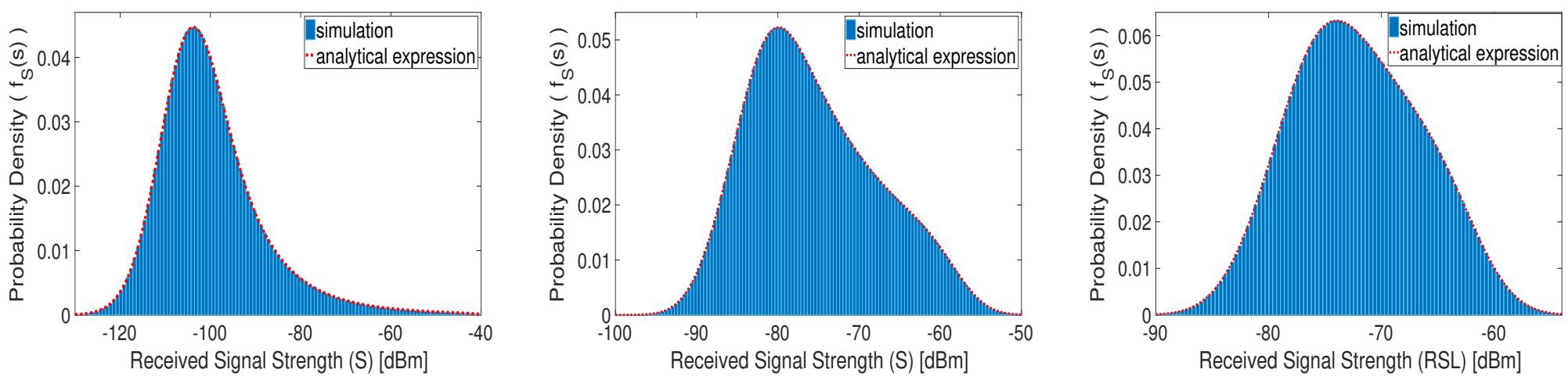

**(a)** $h = 500$ m **(b)** $h = 3500$ m **(c)** $h = 5500$ m

**Fig. 14:** PDF of RSS on ground with changing altitude of UAV for $B = 50^o$

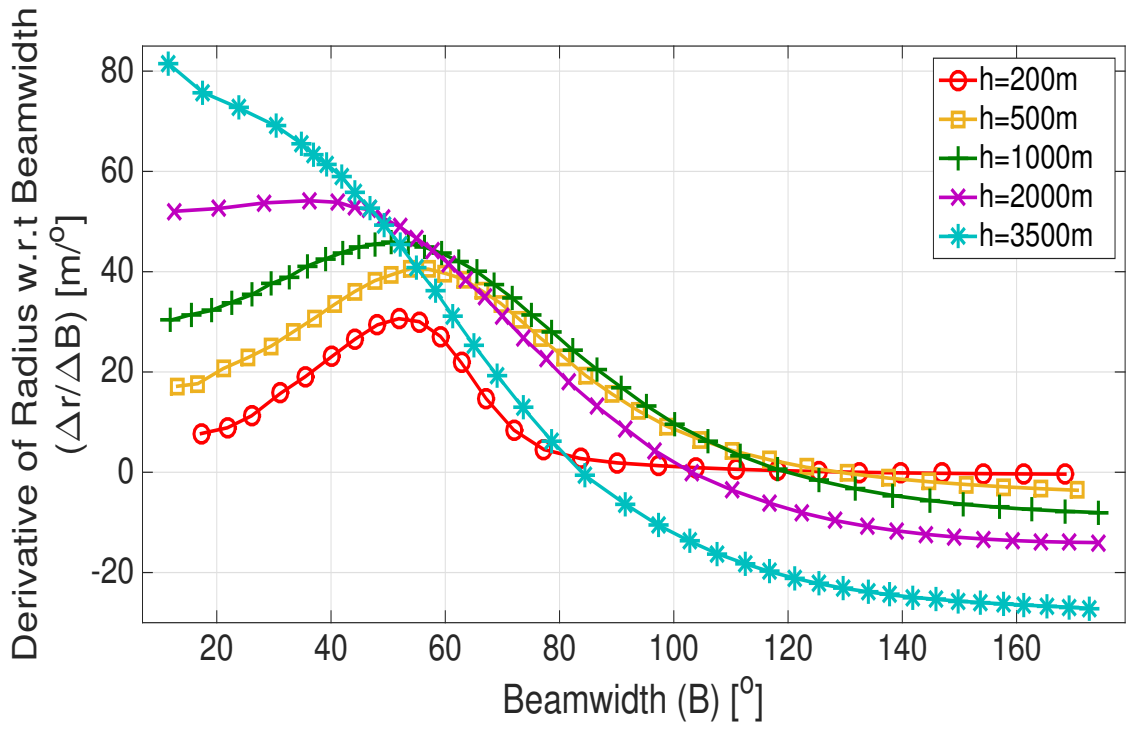

**Fig. 15:** Gradient of coverage radius with respect to beamwidth.

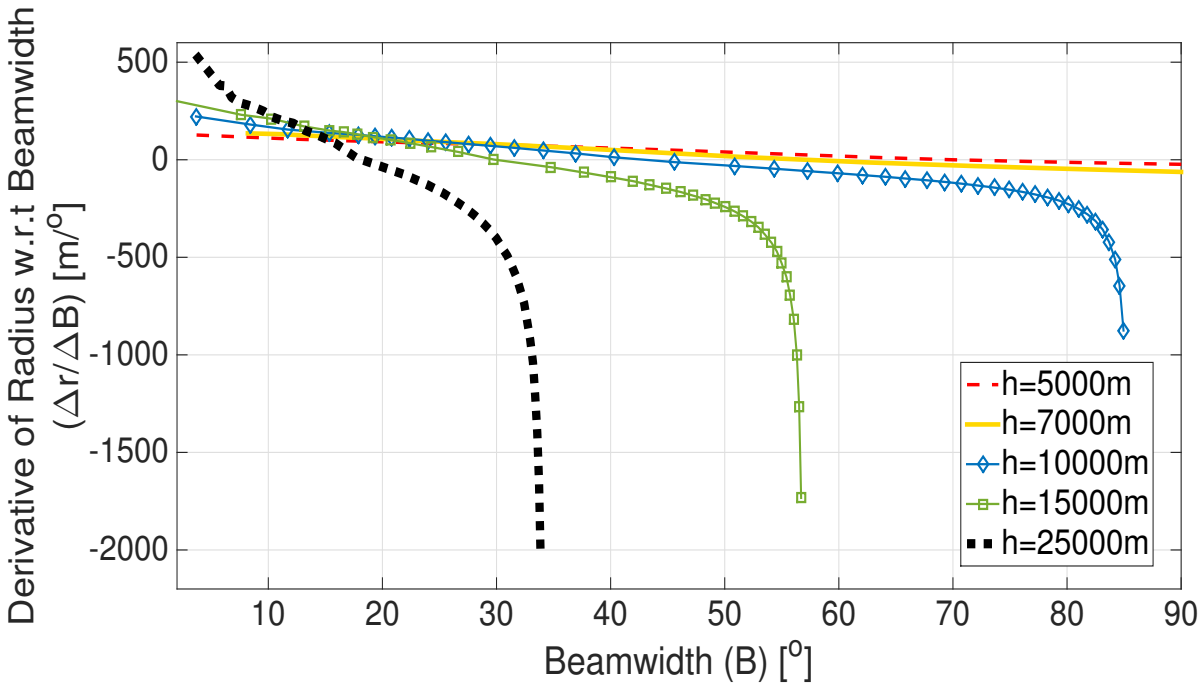

**Fig. 16:** Gradient of coverage radius with respect to beamwidth for different heights in low beamwidth regime.

These figures highlight the capability of our derived equations and the underlying system model to extend the analysis to a wide range of scenarios, such as non-zero tilt angle and asymmetrical beamwidths. Complete analysis of UAV system design in such extended scenarios to provide an even more flexible and on-demand cellular coverage can be focus of a future study.

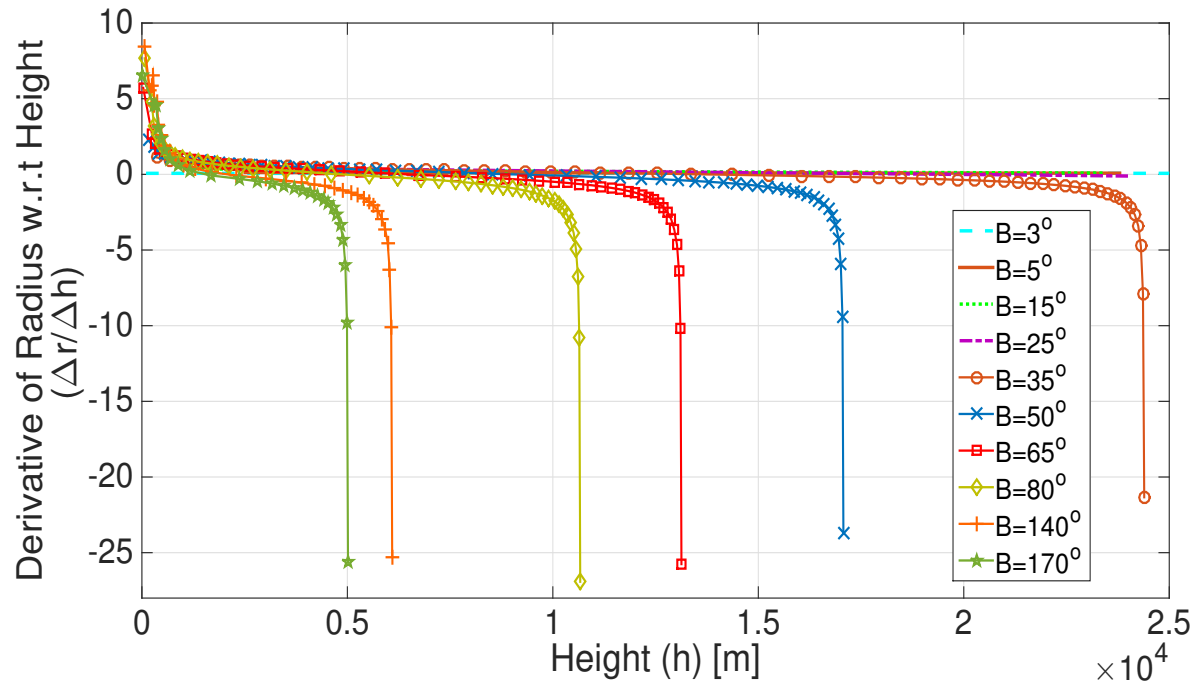

**Fig. 17:** Gradient of coverage radius with respect to height.

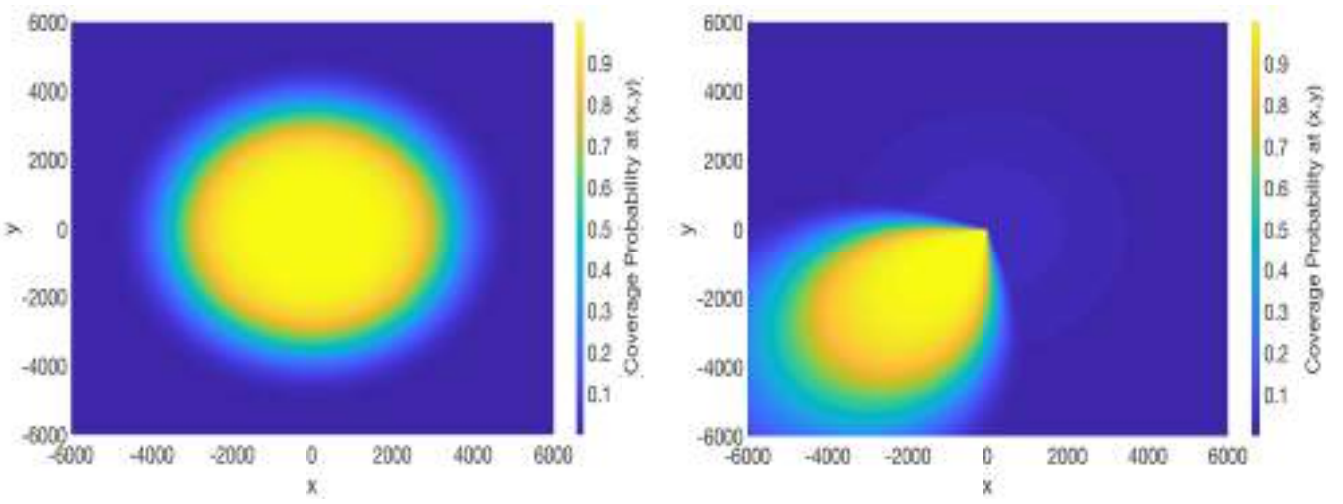

**(a)** $\phi_{tilt} = 0^o$ **(b)** $\phi_{tilt} = 10^o$, $\theta_a = 45^o$

**Fig. 18:** Coverage probability at $B = 50^o$ and $h = 5000$ m.

## G. Coverage Analysis with Multiple UAVs

Previous literature [39], utilizes circle packing theory to determine the number of UAVs to achieve a gain in coverage probability in a certain geographical area. In the circle packing problem, $N$ identical circles (cells) are arranged inside a larger circle (target area) of radius $R_t$ such that the packing density is maximized and none of the circles overlap [46]. The radius of each of the $N$ circles that solves this problem is denoted by $r_{max}$ and one UAV provides coverage to one small cell (cir-

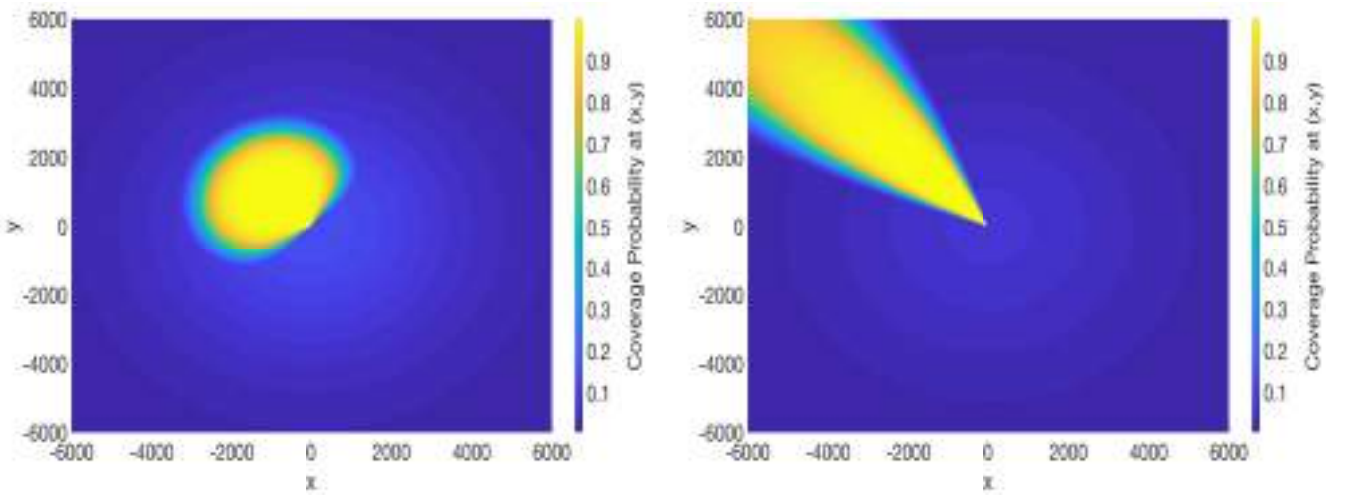

**(a)** $B_\phi = 20^o$, $B_\theta = 70^o$ **(b)** $B_\phi = 70^o$, $B_\theta = 20^o$

**Fig. 19:** Coverage probability with asymmetrical beamwidths at $\phi_{tilt} = 10^o$, $\theta_a = -45^o$and $h = 5000$ m.

**TABLE IV:** Comparison between circular and hexagonal packing in terms of coverage radius of each UAV and maximum total coverage.

| $N$ | $r^c_{max}$ from [39] | $r^h_{max}$ | $C_c$ from [39] | $C_h$ |
|---|---|---|---|---|
| 1 | $R_t$ | $R_t$ | 1 | 1 |
| 2 | 0.500$R_t$ | 0.447$R_t$ | 50.0 | 75.0 |
| 3 | 0.464$R_t$ | 0.500$R_t$ | 64.6 | 75.0 |
| 4 | 0.413$R_t$ | 0.400$R_t$ | 68.6 | 75.0 |
| 5 | 0.370$R_t$ | 0.333$R_t$ | 68.5 | 75.0 |
| 6 | 0.333$R_t$ | 0.286$R_t$ | 66.6 | 75.0 |
| 7 | 0.333$R_t$ | 0.333$R_t$ | 77.8 | 77.8 |
| 8 | 0.302$R_t$ | 0.286$R_t$ | 73.3 | 72.7 |
| 9 | 0.275$R_t$ | 0.250$R_t$ | 68.9 | 69.2 |
| 10 | 0.261$R_t$ | 0.286$R_t$ | 68.7 | 75.0 |

cle). However, this approach towards determining the needed number of UAVs for achieving a coverage level, has two major drawbacks. Firstly, significant gaps between circles or cells are inevitable when two or more circles are used to cover a given area. This is due to the inherent nature of circle packing theory, since in order to cover the target area completely, $N \to \infty$ and $r_{max} \to 0$. Secondly, the number of circles (UAVs) increase rapidly with desired coverage probability. We overcome both of these problems by introducing a UAV placement model that simply uses hexagonal cell shapes instead of circle. This approach not only resolves aforementioned problems but also leads to a better coverage. To illustrate our approach, consider the case for $N = 3$ in Fig. 21. If we consider a hexagonal area with the longest distance from center to the edge denoted by $R_t$ and the distance from center to the vertex of a hexagon by $r_h$, then the maximum distance between any two farthest hexagons will be $4r_h$ as shown in Fig. 21b. Our goal is to minimize this distance in order to maximize the packing density, thus leading to the arrangement shown in Fig. 21c. Here, the maximum distance between farthest hexagons is $4r_h\sqrt{3}/2 < 4r_h$, leading to $r_h = 0.5R_t = r_{max}$. The maximum total coverage (in percentage) for $N = 3$ can then be calculated as follows:

$$C_h = \frac{\text{area covered by 3 UAVs}}{\text{total area to be covered}} = \frac{3\left(\frac{3}{2}\sqrt{3}\left(\frac{1}{2}R_t\right)^2\right)}{\frac{3}{2}\sqrt{3}R_t^2} = 75\% \tag{21}$$

Similar analysis is done for $N = 1$ to 10 and presented in Table IV, where $C_c$ and $C_h$ represents the maximum percentage of actual area covered out of total target area using circular and hexagonal packing respectively while $r^c_{max}$ and $r^h_{max}$ represent the maximum possible radius of each cell using the two approaches. We compare the results of our proposed approach with those from [39] that utilizes circle packing to the same effect. From our proposed approach, the minimum possible coverage is ~70 % for any number of UAVs, whereas with circle packing theory it drops to as low as 50% [39]. This has a direct impact on the coverage threshold requirement of the system. It is highlighted in [39] that a 0.7 coverage performance is impossible to achieve with $1 < N < 7$ using circle packing approach. Our proposed hexagonal packing strategy, on the other hand, ensures that this coverage performance demand can be met with much smaller number of UAVs. We illustrate this by calculating the minimum number of UAVs required to cover different geographical areas by utilizing Table IV for a coverage threshold $\geq$ 70%, with a tolerance of $\pm 1\%$ for $\epsilon = 0.8$ over $0 < h \leq 5000$ and $1 < B < 180$. Fig. 22 compares the resulting minimum number of UAVs obtained with hexagonal packing and circle packing from [39]. For $C = 70\%$, from circle packing approach, we can cover a desired area upto 14 km with 1, 7 or 8 UAVs. On the other hand, with our proposed hexagonal packing approach, we can cover an area with a much smaller number of UAVs, i.e., 1, 2, 4 or 5 UAVs. However, the number of UAVs required to serve multiple users would also depend on transmit power and bandwidth allocation of multiple UAVs. Such considerations are not a focus of this work. The reader is referred to three excellent works which deal with maximizing the minimum average rate of users via joint bandwidth and transmit power allocation [13],[31],[36] for detailed insight into these considerations.

Another advantage that hexagonal packing offers is the relative scalability of number of UAVs as the coverage threshold changes, which in case of circle packing increases rapidly as coverage threshold goes from 50% to 80% as seen in Fig. 22.

Next, we investigate the relationship between number of UAVs and beamwidth of each UAV. First, we find the maximum possible $r$ for a given geographical area with radius $R_t$ using Table IV as the number of UAVs vary and then the corresponding beamwidth using Fig. 3 from our proposed model. Fig. 23 illustrates the results for a UAV deployed at a height of 1000 m to cover a target geographical area of $R_t = 3500$ m in a suburban environment. The overall decreasing trend between beamwidth and number of UAVs for different coverage requirements quantifies the intuitive observation that we can either cover the same area with a single UAV having a wide beamwidth or with multiple UAVs having narrow beamwidths. For example, for a coverage threshold of 60%, a target area of radius 3500 m can be covered either with 10 UAVs, each having a beamwidth of $22^o$ or with a single UAV having a beamwidth of $150^o$. Thus proposed model enables more design options for a wireless system designer with regards to conservation of infrastructure.

Finally, in order to observe the trend of UAV altitude as the number of UAVs vary, the optimal UAV altitude that yields maximum possible coverage is plotted in Fig. 24 for different number of UAVs. The system model using binary antenna gain pattern proposed in earlier studies such as [39] does not reflect

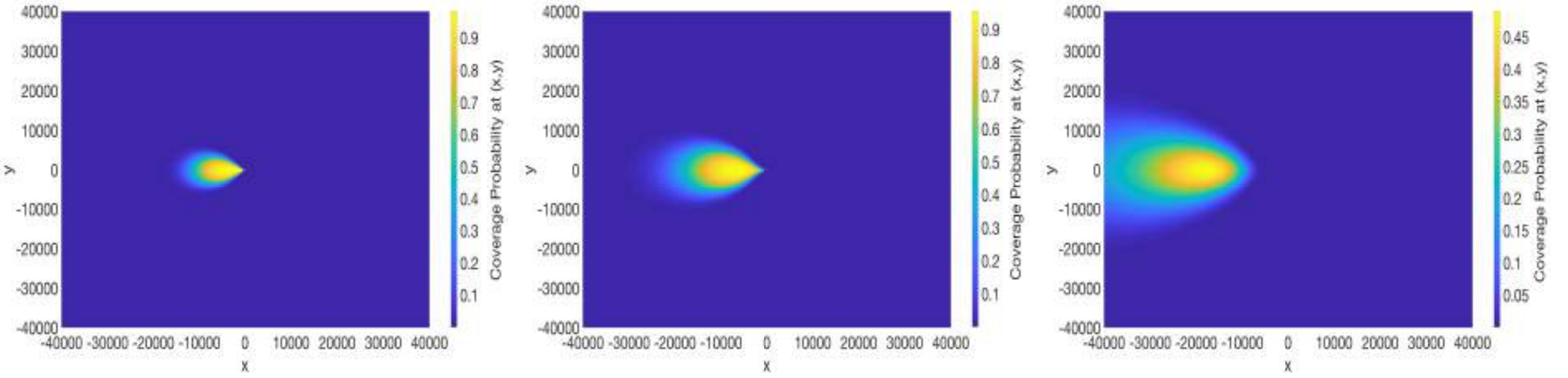

**(a)** $\phi_{tilt} = 20^o$ **(b)** $\phi_{tilt} = 40^o$ **(c)** $\phi_{tilt} = 80^o$

**Fig. 20:** Coverage probability with varying tilt angles at $B = 40^o$ and $h = 10000$ m.

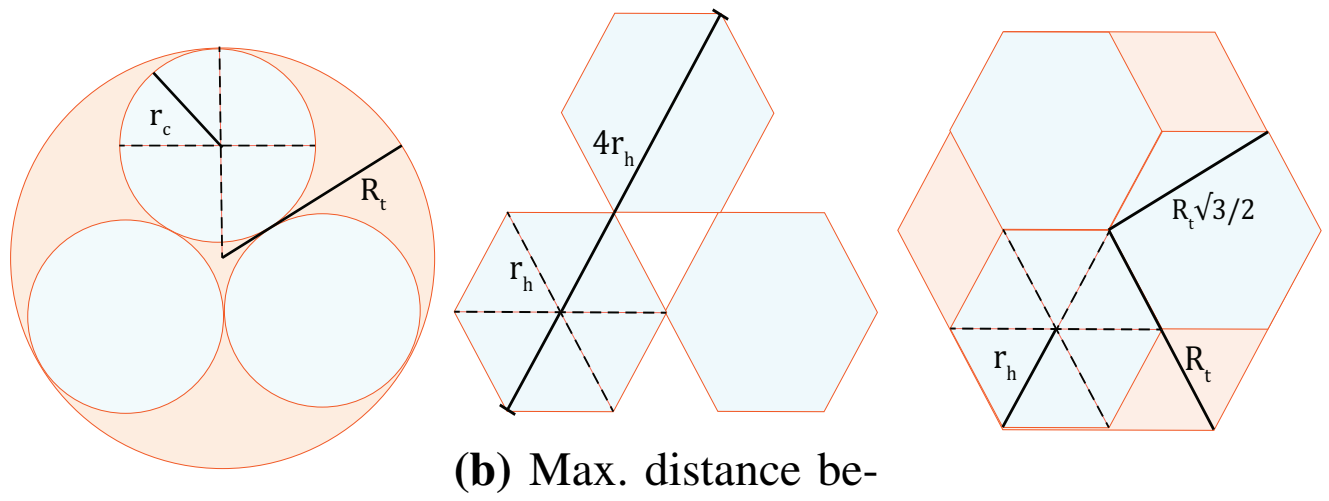

**(a)** Circle packing **(b)** Max. distance between hexagons **(c)** Hexagon packing the role of beamwidth in UAV altitude with increasing number of UAVs. Hence, UAV altitude decreases monotonically as number of UAVs increases. Fig. 24 shows this is not the case when a practical antenna gain pattern, as proposed in this study, is used. In the plot, one UAV covers maximum possible area for a certain beamwidth that can be found from Fig. 4. For beamwidths upto $15^o$, UAV altitude with number of UAVs follow the same trend as in [39]. However, the trend changes for higher beamwidths. This trend is explicitly compared with the model in [39] for a beamwidth of $100^o$. This concludes that, contrary to observation made in prior studies with simple or no antenna models, it is not necessary for UAV altitude to decrease monotonically as the number of UAV increases; in fact, it can also either increase monotonically or behave as a combination of increasing and decreasing altitude as the number of UAVs increase.

**Fig. 21:** Circle packing vs hexagonal packing for $N = 3$.

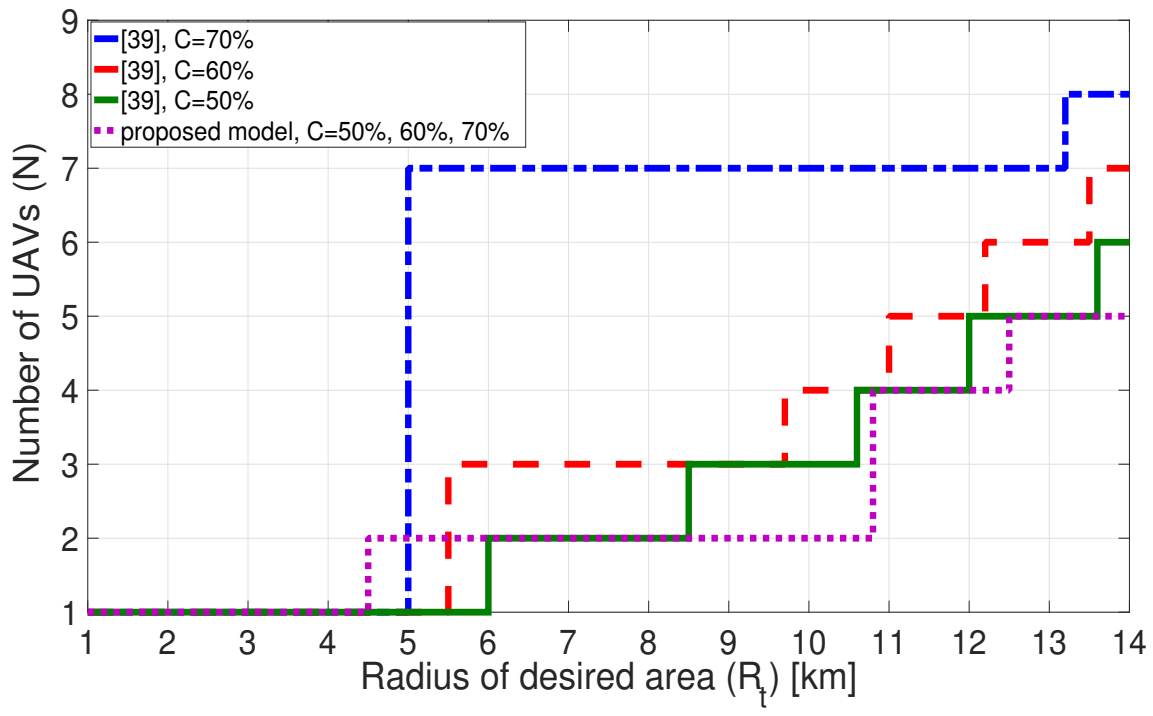

**Fig. 22:** Minimum number of UAVs versus radius of desired area for different minimum coverage thresholds.

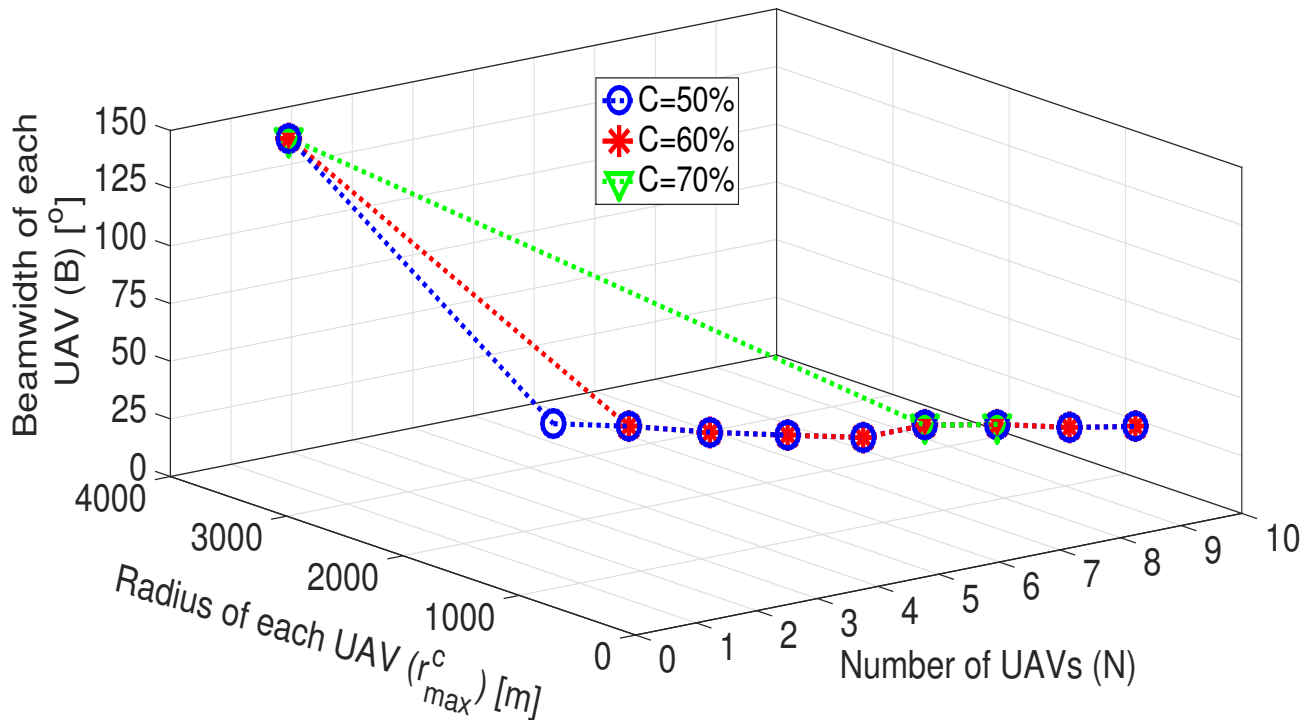

**Fig. 23:** Optimal beamwidth for multiple UAVs and corresponding radius of each UAV.

Presented analysis can also be exploited for interference management in the presence of other aerial platforms since it provides multiple altitude options for UAV deployment, thus leading to more flexible design options which is imperative to the design of next generation cellular systems. Such investigations of interference using proposed model can be focus of a future study. Note that in order to provide full coverage of the considered area, significant overlaps will be inevitable. Several techniques can be leveraged to optimize the number of UAVs needed for full coverage, for example, [47] minimizes the number of UAVs needed to provide wireless coverage for a group of distributed ground terminals, ensuring that each ground termination is within the communication range of at least one UAV. The work in [47] analyzes the UAV placement problem under LoS conditions without antenna model considerations and can be extended to incorporate other elevation angle dependent factors considered in this study. Determining the optimal overlap can be handled using techniques such as adaptive bandwidth allocation among the UAVs as proposed in future work of [36].

For discussions related to other UAV deployment challenges, such as battery considerations, limited payload capacity, security and hostile weather conditions, the reader is referred to [2], [48].

## V. Conclusion

This paper provides a holistic analysis of the interplay between key UAV deployment parameters: coverage radius, height and beamwidth while considering design space dimensions that remain unexplored in existing studies. It further provides a mathematical model to estimate RSS at any distance from boresight of antenna as a function of antenna beamwidth and altitude. The analysis and results provides several new insights that prior models with no or simplified antenna, path loss, or shadowing models do not reveal, such as: 1) UAV altitude or antenna beamwidth does not have to necessarily increase continuously for higher coverage radius; 2) contrary to findings reported in some prior studies, UAV coverage radius does not necessarily increase as altitude increases; 3) the minimum number of UAVs required to cover a given area does not necessarily decrease monotonically as UAV altitude increases. These results allow us to determine optimal UAV parameters for realistic deployment.

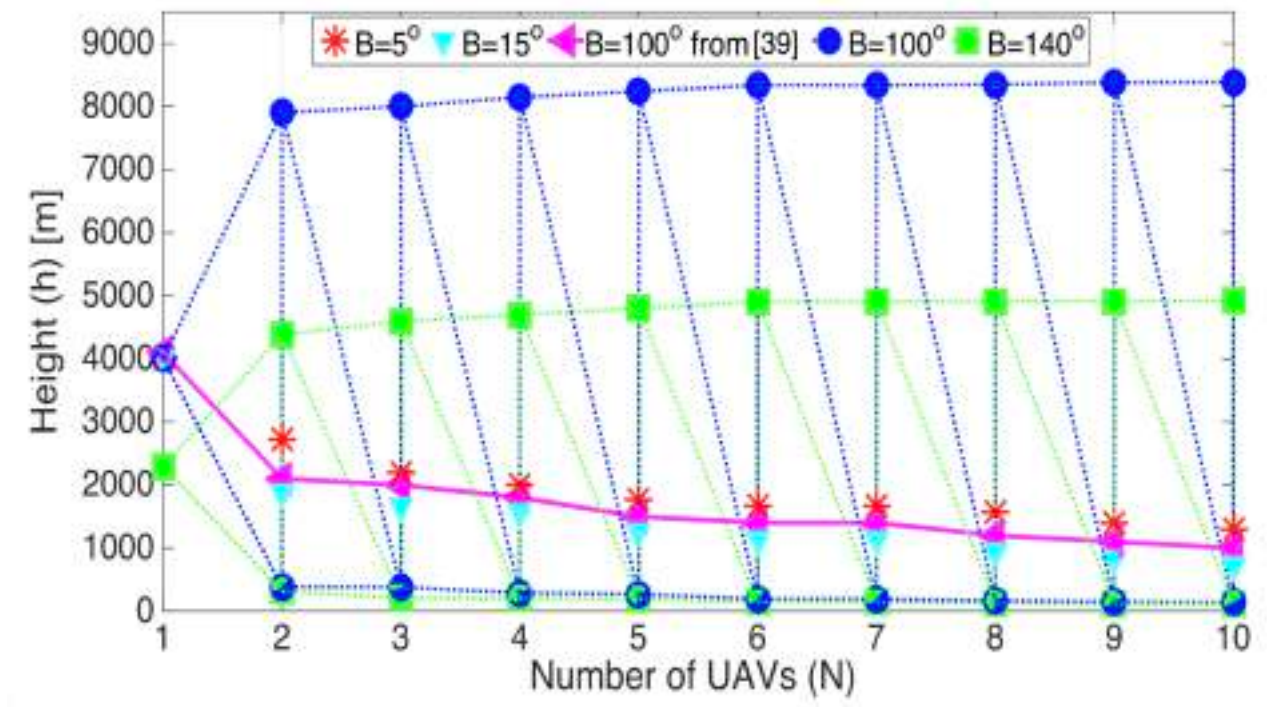

**Fig. 24:** Altitude with varying no. of UAVs for different beamwidths.

Furthermore, based on the analysis of effect of beamwidth and altitude on coverage radius, it is found that antenna beamwidth and altitude should be optimized simultaneously rather than independently as is the case assumed in previous works. It is also concluded that optimizing beamwidth instead of height to control coverage may be a more practical and that coverage is most sensitive to beamwidths of less than $40^o$.

A hexagonal packing is proposed for solving coverage optimization problem with multiple UAVs. The advantage of proposed scheme is that it leaves much smaller coverage holes. Thus it can cover a higher proportion of the given area with same number of UAVs compared to circle packing and is scalable in terms of number of UAVs with increasing probability of coverage.

## Appendix

The probability density function of RSS at an arbitrary point in a cell is found by first deriving PDFs of received signal strengths in LoS and NLoS scenarios using transformations of RVs. Thereafter, the theorem for finding PDF of sums of independent random variables from [49] is applied.

For the scenario with circular coverage pattern of the UAV, we can express the RV, $S_l(h,r,B)$ , from (17) as:

$$\begin{aligned} S_l(h,r,B) &= P_l(\phi_{MS})\, R_l(h,r,B) \\ &= P_l(\phi_{MS})\Bigg[10\log\left(\frac{29000}{B^2}\right) - 12\left(\frac{\tan^{-1}\left(\frac{r}{h}\right)-\phi_{tilt}}{B}\right)^2 \\ &\qquad + T - 20\log\left(\frac{4\pi f d}{c}\right)\Bigg] - P_l(\phi_{MS})\, X_l \\ &= P_l(\phi_{MS})\, A_1 - P_l(\phi_{MS})\, X_l \end{aligned} \tag{22}$$

where $P_l$ is given in (2) and $A_1$ can be treated as a constant for a UAV deployed at a fixed height and beamwidth, given by:

$$A_1 = T + 10\log\left(\frac{29000}{B^2}\right) - 12\left(\frac{\tan^{-1}\left(\frac{r}{h}\right)-\phi_{tilt}}{B}\right)^2 - 20\log\left(\frac{4\pi f d}{c}\right) \tag{23}$$

We proceed by first finding PDF of $S_l$ by applying transformations of random variables as follows:

$$\begin{aligned} F_{S_l}(s_l) &= \mathrm{P}\left(S_l \le s_l\right) \\ &= \mathrm{P}\left(P_l(\phi_{MS})\, A_1 - P_l(\phi_{MS})\, X_l \le s_l\right) \\ &= \mathrm{P}\left(X_l \le \frac{s_l - P_l(\phi_{MS})\, A_1}{P_l(\phi_{MS})}\right) \\ F_{S_l}(s_l) &= F_{X_l}\left(\frac{s_l - P_l(\phi_{MS})\, A_1}{P_l(\phi_{MS})}\right) \end{aligned} \tag{24}$$

where $F_{S_l}(s_l)$ and $F_{X_l}(x_l)$ are the cumulative distribution functions (CDFs) of $S_l$ and $X_l$ respectively. Note that $P_l$ is a function of $\phi_{MS}$, which is in turn a function of $h$ and $r$. However, for compactness, this dependency is omitted in subsequent analysis.

Both sides of (24) are a function of $s_l$ and therefore, we differentiate both sides w.r.t $s_l$ in order to get the PDF:

$$\begin{aligned} f_{S_l}(s_l) &= f_{X_l}\left(\frac{s_l - P_l\, A_1}{P_l}\right)\frac{d}{d\, s_l}\left(\frac{s_l - P_l\, A_1}{P_l}\right) \\ f_{S_l}(s_l) &= \frac{1}{P_l} f_{X_l}\left(\frac{s_l - P_l\, A_1}{P_l}\right) \end{aligned} \tag{25}$$

This allows us to find the PDF of $S_l$ based on the PDF of $X_l$ which is $\mathcal{N}(0,\sigma_n)$ random variable. Applying the transformation in (25) yields the following expression for $f_{S_l}(s_l)$:

$$f_{S_l}(s_l) = \frac{\exp\left(-\frac{(s_l - P_l\, A_1)^2}{2(P_l\, \sigma_l)^2}\right)}{\sqrt{2\pi} P_l\, \sigma_l} \tag{26}$$

Following a similar procedure, we then derive PDF of $S_n$, which yields following expression:

$$f_{S_n}(s_n) = \frac{\exp\left(-\frac{(s_n - [P_n(A_1 - \mu_{sh})])^2}{2(P_n)^2\left(\sigma_n^2 + \sigma_{sh}^2\right)}\right)}{\sqrt{2\pi} P_n \sqrt{\sigma_n^2 + \sigma_{sh}^2}} \tag{27}$$

We can now proceed to derive PDF of $S_r$ by performing convolution of (26) with (27) as in (28), where $A_2$ in (28) is expanded in (29).

Next, we perform algebraic manipulation on (29) in (30) to convert the terms in it to $(C+D)^2$ form in order to apply completing the squares method. In order to complete the square of (30), we define a new variable in the following manner:

$$\sigma_s = \sqrt{\sigma_{s_l}^2 + \sigma_{s_n}^2} \tag{31}$$

where $\sigma_{s_l}^2 = (P_l\, \sigma_l)^2$ and $\sigma_{s_n}^2 = (P_n)^2\left(\sigma_n^2 + \sigma_{sh}^2\right)$.

Therefore, our PDF expression in (28) reduces to:

$$f_{S_r}(s_r) = \int_{-\infty}^{\infty} \frac{1}{\sqrt{2\pi}\sigma_s} \frac{1}{\sqrt{2\pi}\frac{\sigma_{s_l}\,\sigma_{s_n}}{\sigma_s}} \exp\left[-\frac{A_3}{2(\frac{\sigma_{s_l}\,\sigma_{s_n}}{\sigma_s})^2}\right] \tag{32}$$

where $A_3$ equals to (33).

Squares are now completed by collecting the appropriate terms as in (34). Finally, the exponent in (32) can be broken

$$f_{S_r}(s_r) = \int_{-\infty}^{\infty} \frac{1}{\sqrt{2\pi}\,P_n\sqrt{\sigma_n^2+\sigma_{sh}^2}} \exp\left[-\frac{(s_r - s_l - [P_n\,(A_1-\mu_{sh})])^2}{2\,(P_n)^2\,(\sigma_n^2+\sigma_{sh}^2)}\right] \frac{1}{\sqrt{2\pi}\,P_l\,\sigma_l} \exp\left[-\frac{(s_l - P_l\,A_1)^2}{2\,(P_l\,\sigma_l)^2}\right] d\,s_l$$
$$= \int_{-\infty}^{\infty} \frac{1}{\sqrt{2\pi}\sqrt{2\pi}\,P_n\sqrt{\sigma_n^2+\sigma_{sh}^2}\,P_l\,\sigma_l} \exp\left[-\frac{A_2}{2\,(P_l)^2\,\sigma_l^2\,(P_n)^2\,(\sigma_n^2+\sigma_{sh}^2)}\right] d\,s_l \quad (28)$$

$$A_2 = (P_l)^2\,\sigma_l^2\,(s_r - s_l - [P_n\,(A_1-\mu_{sh})])^2 + (P_n)^2\left(\sigma_n^2+\sigma_{sh}^2\right)(s_l - P_l\,A_1)^2$$
$$= (P_l)^2\,\sigma_l^2\,\left(s_r^2 + s_l^2 + [P_n\,(A_1-\mu_{sh})]^2 - 2s_l\,s_r - 2s_r\,[P_n\,(A_1-\mu_{sh})] + 2s_l[P_l(A_1-\mu_{sh})]\right) +$$
$$(P_n)^2\left(\sigma_n^2+\sigma_{sh}^2\right)\left(s_l^2 + (P_l\,A_1)^2 - 2s_l\,P_l\,A_1\right) \quad (29)$$

$$A_2 = s_l^2\left[(P_l)^2\,\sigma_l^2 + (P_n)^2\left(\sigma_n^2+\sigma_{sh}^2\right)\right] - 2s_l\left[(P_l\,\sigma_l)^2\,(s_r - P_n(A_1-\mu_{sh})) + (P_n)^2\left(\sigma_n^2+\sigma_{sh}^2\right)P_l\,A_1\right] +$$
$$(P_l\,\sigma_l)^2\left(s_r^2 + [P_n\,(A_1-\mu_{sh})]^2 - 2s_r\,[P_n\,(A_1-\mu_{sh})]\right) + (P_n)^2\left(\sigma_n^2+\sigma_{sh}^2\right)[P_l\,A_1]^2 \quad (30)$$

$$A_3 = s_l^2 - 2s_l\frac{\sigma_{s_l}^2(s_r - [P_n\,(A_1-\mu_{sh})]) + \sigma_{s_n}^2 P_l A_1}{\sigma_s^2} + \frac{\sigma_{s_l}^2(s_r^2 + [P_n\,(A_1-\mu_{sh})]^2 - 2s_r\,[P_n\,(A_1-\mu_{sh})] + \sigma_{s_n}\,[P_l\,A_1]^2}{\sigma_s^2} \quad (33)$$

$$A_3 = \left(s_l - \frac{\sigma_{s_l}^2\,(s_r - [P_n\,(A_1-\mu_{sh})]) + \sigma_{s_n}^2 P_l\,A_1}{\sigma_s^2}\right)^2 - \left(\frac{\sigma_{s_l}^2\,(s_r - [P_n\,(A_1-\mu_{sh})]) + \sigma_{s_n}^2 P_l\,A_1}{\sigma_s^2}\right)^2 + \left(\frac{\sigma_{s_l}^2\,(s_r - [P_n\,(A_1-\mu_{sh})])^2 + \sigma_{s_n}^2 P_l^2\,A_1^2}{\sigma_s^2}\right) \quad (34)$$

$$f_{S_r}(s_r) = \frac{1}{\sqrt{2\pi\sigma_s}} \exp\left[-\frac{(s_r - [P_l\,A_1 + P_n\,(A_1-\mu_{sh})])^2}{2\sigma_s^2}\right] \underbrace{\int_{-\infty}^{\infty} \frac{1}{\sqrt{2\pi\frac{\sigma_{s_l}\sigma_{s_n}}{\sigma_s}}} \exp\left[-\frac{s_l - \frac{\sigma_{s_l}^2(s_r - [P_n(A_1-\mu_{sh})]) + \sigma_{s_n}^2[P_l\,A_1]}{\sigma_s^2}}{2(\frac{\sigma_{s_l}\sigma_{s_n}}{\sigma_s})^2}\right] d\,s_l}_{=1} \quad (35)$$

into a product of two exponents as shown in (35). By noting that the integral in (35) is in fact a Gaussian distribution on $S_l$ (and hence integrates to 1), substituting $\sigma_s$ from (31), $P_l$ and $P_n$ from (2) leads to the expression of PDF of $S_r$ in (18).

## Acknowledgment

This material is based upon work supported by the National Science Foundation under Grant Numbers 1619346, 1559483, 1718956 and 1730650. The statements made herein are solely the responsibility of the authors. For more details, please visit: http://www.ai4networks.com.

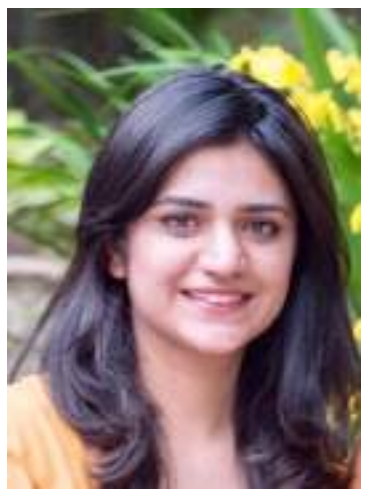

**Haneya Naeem Qureshi** Haneya Naeem Qureshi (GSM'16) received her BS degree in Electrical Engineering from Lahore University of Management Sciences (LUMS), Pakistan, in 2016 and M.S. degree in Electrical and Computer Engineering from the University of Oklahoma, USA in 2017. She is currently pursuing the Ph.D. degree in Electrical and Computer engineering with the University of Oklahoma, USA working in the Artificial Intelligence (AI) for Networks Laboratory, where she is contributing to several NSF-funded projects. Her current research interests include network automation and combination of machine learning and analytics for future cellular systems. She has also been engaged in addressing channel estimation and pilot contamination problem in Massive MIMO TDD systems.

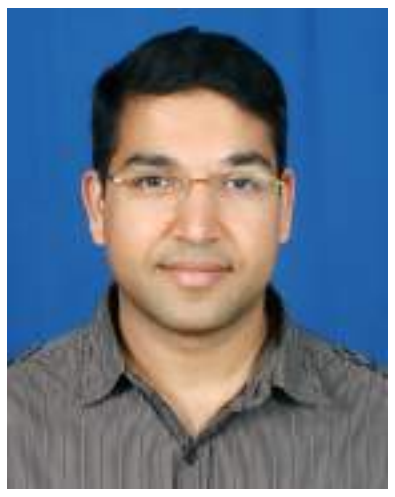

**Ali Imran** Ali Imran (M'15) received the B.Sc. degree in electrical engineering from the University of Engineering and Technology, Lahore, Pakistan, in 2005 and the M.Sc. degree (with Distinction) in mobile and satellite communications and the Ph.D. degree from the University of Surrey, Guildford, U.K., in 2007 and 2011, respectively. He is an Assistant Professor in telecommunications with the University of Oklahoma, Tulsa, OK, USA, where he is the Founding Director of the Artificial Intelligence (AI) for Networks Laboratory (AI4Networks) Research Center and TurboRAN 5G Testbed. He has been leading several multinational projects on Self Organizing Cellular Networks such as QSON, for which he has secured research grants of over 3 million in last four years as the Lead Principal Investigator. He is currently leading four NSF funded Projects on 5G amounting to over 2.2 million. He has authored over 60 peer-reviewed articles and presented a number of tutorials at international forums, such as the IEEE International Conference on Communications, the IEEE Wireless Communications and Networking Conference, the European Wireless Conference, and the International Conference on Cognitive Radio Oriented Wireless Networks, on his topics of interest. His research interests include self-organizing networks, radio resource management, and big-data analytics. He is an Associate Fellow of the Higher Education Academy, U.K., and a member of the Advisory Board to the Special Technical Community on Big Data of the IEEE Computer Society.